\documentclass[11pt]{article}
  \usepackage{amsthm}
  \usepackage{amssymb}
  \usepackage{amsmath}
  \usepackage{mathrsfs}
  \usepackage[all]{xy}
  \usepackage{bbm}
  \usepackage{geometry}
\newcommand{\cA}{\mathcal{A}}

\newcommand{\cC}{\mathcal{C}}

\newcommand{\cH}{\mathcal{H}}

\newcommand{\cO}{\mathcal{O}}
\newcommand{\cP}{\mathcal{P}}

\newcommand{\cR}{\mathcal{R}}

\newcommand{\cZ}{\mathcal{Z}}

\newcommand{\rC}{\mathrm{C}}

\newcommand{\rI}{\mathrm{I}}

\newcommand{\rS}{\mathrm{S}}

\newcommand{\rZ}{\mathrm{Z}}

\newcommand{\sA}{\mathscr{A}}

\newcommand{\sR}{\mathscr{R}}

\newcommand{\Si}{\Sigma} 

\newcommand{\A}{\mathcal{A}} 
\newcommand{\Al}{\mathscr{A}} 
\newcommand{\Bh}{\mathfrak{B}(\mathcal{H})} 

\newcommand{\si}{\sigma}

\newcommand{\eps}{\varepsilon}
\newcommand{\io}{\iota}

\newcommand{\la}{\lambda}

\newcommand{\sst}[1]{\scriptscriptstyle{#1}}

\newcommand {\norm}[1]{\Vert{#1}\Vert}    
\newcommand {\con}[1]{\overline{{#1}}}
\newcommand {\dec}[1]{{\langle{#1}\rangle}}
\newcommand{\minperp}{{\scriptscriptstyle{\perp}}}
\newcommand{\R}{\mathbb{R}} 
\newcommand{\Hil}{\mathcal{H}} 
\newcommand{\Loops}[1]{\mathrm{Loops}_{#1}}

 \author{\null\\Romeo Brunetti$^{1}$ and Giuseppe Ruzzi$^{2}$ \\
\null\\
  \null\\
  \small{$^{1}$Dipartimento di Matematica, Facolt\`a di Scienze MFN,
Universit\`a  di Trento,}\\
    \small{Via Sommarive 14, I-38050 Povo (TN), Italy}\\[5pt] 
  \small{$^{2}$Dipartimento di Matematica, 
                 Universit\`a di Roma ``Tor Vergata,'' }\\
     \small{Via della Ricerca Scientifica, I-00133 Roma,  Italy}  \\[5pt]
           \small{\texttt{brunetti@science.unitn.it,  ruzzi@mat.uniroma2.it}}\\[20pt]
           {\bf\small Dedicated to Klaus Fredenhagen on the occasion of his sixtieth birthday}}

  \title{Quantum charges and spacetime topology: \\
     The emergence of new superselection sectors}

\date{}

\begin{document}

 \maketitle

\begin{abstract}
In which is developed a new form of superselection sectors of 
topological origin. 
By that it is meant a new investigation that includes several extensions of the traditional framework of Doplicher, 
Haag and Roberts in local quantum theories. At first we generalize the notion of representations of nets of 
$\rC^*$--algebras, then we provide a brand new view on selection criteria by adopting one with a strong 
topological flavour. We prove that it is coherent with the older point of view, hence a clue to a genuine extension. 
In this light, we extend Roberts' cohomological analysis to the case 
where 1--cocycles bear non trivial unitary 
representations of the fundamental group of the spacetime, equivalently of its Cauchy surface in case of 
global hyperbolicity. A crucial tool is a notion of group von Neumann 
algebras generated by the 1--cocycles 
evaluated on loops over fixed regions. One proves that these group von Neumann algebras are localized at the bounded region where loops 
start and end and to be factorial of finite type $I$. All that amounts to a new invariant, in a topological sense, 
which can be defined as the dimension of the factor. We prove that any 1--cocycle can be factorized into a part that 
contains only the charge content and another where only the topological information is stored. 
This second part resembles much what in literature are known as geometric phases. 
Indeed, by the very geometrical origin of the 1--cocycles that we discuss in the paper, 
they are essential tools in the theory of net bundles, and the topological part is related 
to their holonomy content. At the end we prove the existence of net representations. 
\end{abstract}


  \theoremstyle{plain}
  \newtheorem{definition}{Definition}[section]
  \newtheorem{theorem}[definition]{Theorem}
  \newtheorem{proposition}[definition]{Proposition}
  \newtheorem{corollary}[definition]{Corollary}
  \newtheorem{lemma}[definition]{Lemma}
  
  \theoremstyle{definition}
  \newtheorem{remark}[definition]{Remark}
  \newtheorem{example}[definition]{Example}

  \theoremstyle{plain}
  \newtheorem*{Main}{Main Theorem}
  \newtheorem*{MainT}{Main Technical Theorem}

 \newpage


\tableofcontents
\markboth{Contents}{Contents}
\newpage


\numberwithin{equation}{section}

\section{Introduction}
A large class of effects in physics can be explained using the features and language of topology. 
Starting from those due to non-trivial topology of configuration spaces in classical and quantum physics, 
one also finds intriguing topological effects in general relativity
and quantum field theory. To name just a few of the most interesting
ones we recall the 
Ehrenberg-Siday-Aharonov-Bohm effect \cite{ES,AB} in quantum mechanics and its generalization named Pancharatnam-Berry phase \cite{Pa, Be}, or its classical counterpart called Hannay's angle \cite{Han}, the Jahn-Teller effect \cite{JT}, Wheeler's geons \cite{Wh} and the Casimir effects \cite{Ca}.

The aim of the present paper is to trace the route for a rigorous attack to the dependence on topology of certain structures in quantum field theory. In particular, we aim at a complete model independent description, at least in a preliminary case where the topological effects under interest are those related to specific topological features of spacetimes. The language is that of local quantum theory \cite{Ha, BF2} (otherwise called algebraic quantum field theory) where, as it is well known, one finds the best understanding about the nature and properties of structural properties of quantum field theories. The proper setting is that related to the prominent case of superselection sectors where charged quantum numbers find their definition as attributes associated to unitary equivalence classes of representations of the net of local observable algebras satisfying certain selection conditions.

The traditional analysis of such selection criteria -- and associated equivalence classes of representations thereof -- goes mainly by the names of Doplicher, Haag, and Roberts \cite{DHR}. They  worked out the structure of charges localizable into bounded regions, whilst it is due to Buchholz and Fredenhagen \cite{BF} the study of charges that can be localized in unbounded regions, i.e. spacelike cones. All that was done for quantum field theories on Minkowski spacetime in dimensions $d\ge 3$. Other groups of researchers have been able to follow the same route in various directions, especially in the direction of conformal quantum field theory in two dimensions, and besides the crucial results of Fredenhagen, Rehren and Schroer \cite{FRS}, the main success was obtained by Kawahigashi and Longo in \cite{KL}, where they have been able to completely classify theories with central charge less than one. 

The authors of the present paper have recently put forward an analysis of the structure of superselection sectors 
\cite{BR} that provides a new perspective both by the adopted techniques and in the fact that superselection theory
is now applicable to the larger setting of locally covariant quantum field theories \cite{BFV}. The obtained results 
confirm that sectors of the kind that Doplicher, Haag and Roberts studied long ago find their most natural position in 
the locally covariant framework. In fact, we can associate with any 4-dimensional globally 
hyperbolic spacetime  a unique, symmetric, tensorial $\rC^*$--category (that possesses conjugates in case of finite 
statistics) and that to any isometric embedding between such spacetimes the previous categories can be contravariantly
related as to guarantee that charges are preserved under the embedding. 

The local covariance of sectors comes from the analysis of 1--cocycles associated with the Roberts' cohomology of 
posets \cite{Rob76,Rob80,Rob90,Rob00,Ruz3} that carry a trivial unitary representation of 
the fundamental group  of the spacetime. 

It is natural to try to understand the kind of 1--cocycles that carry
a non-trivial unitary representation and 
to see whether one can associate with them a different kind of superselection sector and charge, now attributable 
to the possible non-trivial topology of the spacetime. 
The main results of the following analysis show that this 
is indeed possible and fruitful. 

We have now a clear relation between topological properties of a 
spacetime and structural properties of 1--cocycles 
carrying non-trivial representations of the fundamental group of the spacetime. 
The analysis that follows, however, is not yet casted into a locally covariant form, 
although our initial aim was into that direction. We will be working on a fixed, but otherwise arbitrary, 
4-dimensional globally hyperbolic spacetime. We hope to return elsewhere to the locally covariant analysis, 
and consider this paper as the third one of the announced series in \cite{BR}.

The main ingredients are the following: first, a generalization of the usual notion of representations 
of a net of local algebras, something that we termed ``unitary net representations;'' secondly, the association 
of this new notion with a 1--cocycle, in the sense of Roberts'
cohomology of posets; thirdly, a new selection criterion that generalizes that of Doplicher, Haag and Roberts, 
for the sake of attributing a non-trivial dependence on the spacetime topology
to the superselection sectors so defined; fourthly, one defines a von Neumann algebra which is the group algebra 
generated by a 1--cocycle evaluated on all loops over a fixed bounded region, as a starting and ending point, 
and proves that this algebra is localized, i.e. it is a subalgebra of the von Neumann algebra of the net that is 
localized in the chosen region. 
This last ingredient is the key structural element of the analysis that follows. It allows us to 
attribute to each 1--cocycle generating its own group von Neumann algebra a new invariant, 
called ``topological dimension,'' that resembles much, and has similar properties of a charge quantum number, 
and carries non-trivial informations about the topology of the spacetime.

Furthermore, we prove that any 1--cocycle can be splitted into a part that carry only information on the charge 
content of the sector and a part that carry the topological information of spacetime. This last resembles much 
what we cited at the beginning as geometric phases. For more on that see also Section \ref{F}, where  we also prove the existence of net representations.

Abstract as they are, one would like to have concrete examples of constructions of this non-Abelian kind of geometric phases.
A recent work, done by one of the authors in collaboration with Franceschini and Moretti \cite{BFM}, 
provides a first explicit example of a 1--cocycle induced by the non-trivial topology of spacetime in 
the simple case of massive bosonic quantum field theory on the 2-dimensional Einstein cylinder.

A further glance at models may indicates other situations where our
analysis may apply. For instance, in cosmology one looks for visible 
effects of the non-trivial topology of spacetimes by searching for 
additional images in the sky of the same galaxy, due to
the possible presence of a cosmic string.
Besides that, we mention also that there is a large class of physically meaningful multiply connected spacetimes.
These spacetimes are a class of Friedmann-Lama\^{\i}tre models, solutions of the Einstein equations which are used as cosmological models (see \cite{LL,Sou}). 

We finally point the reader to a recent interesting paper by Morchio and Strocchi \cite{MS} where, in the case of quantum mechanics on manifolds, they describe a classification of topological effects in close analogy to our results. Also, papers by D\"obner et alt. \cite{Dob} and Landsman \cite{KLa}, have a similar flavour.

The paper has been structured in such a way to mantain a decent ratio between size and completeness, hence many results are presented into the simplest form that we could think of. We refer the reader to \cite{LR,DR89,Ruz3,BR} for a deeper introduction to some of the mathematical notions that we use.

\section{The category of net representations}
\label{A}
We introduce the notion of net representations  for  nets of 
$\rC^*$--algebras. We analyze in particular the class
of unitary net representations pointing out their topological content.  
The importance of this new notion resides 
in the fact that a new class of superselection sectors induced by the topology of spacetimes is described in terms of net representations (see next sections).  We shall use the tool of cohomology of posets 
to make explicit the topological information carried by 
net representations. Within this section we shall also discuss, very briefly, preliminary informations on the simplicial set associated with a  poset and the first degree of its cohomology.  
Details can be found in \cite{Rob00,Ruz3,BR, RRV}.\bigskip

Let $K$ be a poset  with order relation $\leq$.
We consider the simplicial set $\Si_*(K)$ of \emph{singular simplices}
associated with $K$. We use the standard symbols 
$\partial_i$ and $\si_i$ to denote the face and degeneracy maps, and 
denote the compositions $\partial_i\partial_j$, $\si_i\si_j$ 
respectively  by $\partial_{ij}$ and $\si_{ij}$. We pass now to a brief definition of the set $\Si_n(K)$ of $n$--simplices. 
A $0$--simplex is just an element of $K$. Inductively, for $n\geq 1$, 
and $n$--simplex $x$  is formed by $n+1$, $(n-1)$--simplices
$\partial_0x,\ldots, \partial_nx$ and by an element of the poset 
$|x|$, called the \emph{support} of $x$, such that 
$|\partial_ix|\leq |x|$ for $i=0,\ldots,n$. We shall denote 
0--simplices either by $a$ or by $o$, 1--simplices by $b$, and
2--simplices by $c$. Given a 1--simplex $b$ the \emph{reverse} 
$\overline{b}$ is the 1--simplex having the same support as $b$ 
and such that $\partial_0\overline{b}=\partial_1b$,
$\partial_1\overline{b}=\partial_0b$.\smallskip

\indent Composing 1--simplices one gets paths. 
A \emph{path} $p$ is a finite  ordered set of 1--simplices $b_n*\cdots *b_1$
satisfying the relations $\partial_0b_{i-1}= \partial_1b_{i}$ 
for $i=2,\ldots,n$. We define the 0--simplices  
$\partial_1p\doteq\partial_1b_1$ and
$\partial_0p\doteq\partial_0b_n$ and call them, respectively, 
the \emph{starting} and the \emph{ending} point of $p$. The \emph{support} of a path is defined as the union of the supports of the 1--simplices by which it is composed.
By 
$p:a\to \tilde a$ we mean a path starting from $a$
and ending at $\tilde a$. The \emph{reverse} of $p$ 
is the path $\overline{p}:\tilde a\to a$ defined by 
$\overline{p}\doteq \overline{b}_1*\cdots *\overline{b}_n$. 
If $q$ is a path from $\tilde a$ to $\hat
a$, then we can define, in an obvious way, the composition 
$q*p:a\to \hat a$. The poset $K$ is said to be \emph{pathwise
connected} whenever for any pair $a$,$\tilde a$ of 0--simplices
there is a path from $a$ to $\tilde a$.\smallskip

Let $p=b_n*\cdots*b_1$ be a path. An \emph{elementary deformation} 
of $p$ consists in replacing a 1--simplex $\partial_1c$ of the path by
the pair $\partial_0c*\partial_2c$, where $c\in\Si_2(K)$, 
or conversely in replacing a consecutive pair
$\partial_0c*\partial_2c$ by a single 1--simplex $\partial_1c$.
Two paths with the same endpoints are \emph{homotopic} 
if they can be obtained from one another by a finite sequence of
elementary deformations. Homotopy defines an equivalence relation $\sim$ 
on the set of paths with the same endpoints which is compatible 
with reverse and composition. The \emph{first homotopy
group} of the poset $\pi_1(K,a)$, with base point $a$, is the quotient 
of the set $\Loops{K}(a)$ of all paths $p:a\to a$ in $K$ with respect to the homotopy equivalence relation. If $K$ is pathwise connected the first 
homotopy group does not depend, up to isomorphism, 
on the base point. The isomorphism class is  
the fundamental group $\pi_1(K)$  of the poset and we will say 
that $K$ is simply connected whenever the fundamental 
group is trivial.\smallskip

We conclude this introductive part with two remarks. 
First, we recall that if $K$ is upward 
directed, namely if for any 
$a_1,a_2\in K$ there is $a_3\in K$ with $a_1,a_2\leq a_3$, 
then it is pathwise and simply connected. Secondly, 
let $M$ be a topological space and consider a basis of its topology 
formed by open arcwise and simply-connected  open sets of $M$. 
If $K$ is a poset formed by the elements of this basis with 
the inclusion order relation $\subseteq$, 
then $\pi_1(M)=\pi_1(K)$.\bigskip 

We now turn to the definition of net representations. 
From now on we fix a poset $K$ and assume that it is pathwise connected.
A \emph{net} of $\rC^*$--algebras $\cA_K$
on a poset  $K$ is given by the following data: 
there is mapping $a\mapsto\A(a)$ from $K$ 
to unital $\rC^*$--algebras; for any pair $\tilde a, a\in K$ with $\tilde a\leq a$, there is an injective  $*$--morphism    
$\jmath_{a\tilde a}:\A(\tilde a)\to \A(a)$. 
The morphisms $\jmath_{a\tilde a}$ are called \emph{inclusion morphisms}. 
These  morphisms are required to satisfy the following 
coherence property
\begin{equation}
\label{A:1}
\jmath_{a\tilde a}\, \jmath_{\tilde a\hat a} = \jmath_{a\hat a}\ ,
      \qquad 
      \hat a\leq \tilde a\leq a\ .
\end{equation}
A \emph{net representation} of  
$\A_K$  is  a pair $\{\pi,\psi\}$, where $\pi$ denotes 
a function that associates 
a representation $\pi_a$ 
of $\cA(a)$ on a Hilbert space $\cH_a^\pi$ with 
any $a\in K$; $\psi$ denotes a function that associates 
an injective linear operator 
$\psi_{a\tilde a}:\Hil^\pi_{\tilde a}\to \Hil^\pi_a$ with 
any pair $a,\tilde a\in K$, with $\tilde a\leq a$. The functions $\pi$
and $\psi$ are required to satisfy the following relations
\begin{equation}
\label{A:2}
\psi_{a\tilde a}\,\pi_{\tilde a}  = \pi_a\, \jmath_{a\tilde a}\,
      \psi_{a\tilde a}\ ,  \ \ \ \tilde a \leq a\ , \ \ \mbox{ and } \ \ 
\psi_{a\tilde a}\, \psi_{\tilde a\hat a} = \psi_{a\hat a}\ , \ \ \ 
      \hat a\leq \tilde a\leq a\ . 
\end{equation}
An \emph{intertwiner}  from  
$\{\pi,\psi\}$ to  $\{\rho,\phi\}$ is a function $T$  associating 
a bounded linear operator $T_a:\Hil_a^\pi\to \Hil_a^\rho$ with any 
$a\in K$, and satisfying the relations 
\begin{equation}
\label{A:3}
T_a \, \pi_a = \rho_a \, T_a\ ,  \ \ \mbox{ and } \ \ 
  T_a \, \psi_{a\tilde a} = \phi_{a\tilde a}\, T_{\tilde a}\ , \ \ 
     \tilde a \leq a\ . 
\end{equation}
We denote the set of intertwiners from 
$\{\pi,\psi\}$ to   $\{\rho,\phi\}$ by the symbol
$\big(\{\pi,\psi\},\{\rho,\phi\})$, and say that the net 
representations are \emph{unitarily equivalent}  if they have  a unitary 
intertwiner $T$, that is, $T_a$ is a unitary operator 
for any $a$.\smallskip

The definition of net representation is suggested by 
the theory of bundles over posets \cite{RRV}. 
There is, in fact, an underlying structure of 
a net bundle and, as we shall point out, 
some results on net representations are analogous to those of net
bundles. The contact point resides in the defining properties of the 
function $\psi$, which are the same as the net structure of a net bundle. 
However, net representations already appeared 
in the literature of algebraic quantum field theory, 
although not in this general form. 
They have been considered by 
Buchholz, Haag and Roberts in an unpublished 
paper.\footnote{Private communication by J. E. Roberts.}           
Fredenhagen and Haag encountered this class of representations 
in the reconstruction of a theory from its germs \cite{FH}.  
An argument used in that paper allows us to show 
how net representations arise. We call a \emph{net state} 
a function $\omega$ associating a state $\omega_a$ of $\cA(a)$ with any 
$a\in K$, and which is compatible with the inclusion morphisms, i.e.,  
\begin{equation}
\label{A:3a}
\omega_a\,j_{a\tilde a} =\omega_{\tilde a}\ , \qquad \tilde a\leq a\ . 
\end{equation} 
Given a net state $\omega$,
denote  the GNS--construction  
of $\omega_a$ by  $\{\pi_a,\Hil_a,\Omega_a\}$, and define
\begin{equation}
\label{A:3b}
\psi_{a\tilde a}\,\pi_{\tilde a}(A)\,\Omega_{\tilde a} \doteq \pi_{a}(\jmath_{a\tilde a}(A))\,\Omega_a\ , \qquad \tilde a\leq a\ .
\end{equation}
By using (\ref{A:3a}), we have that 
$\psi_{a\tilde a}: \cH_{\tilde a}\to \cH_a$ 
is an isometry. Furthermore, it is a routine calculation to check 
that $\psi_{a\tilde a}\pi_{\tilde a} = \pi_a\jmath_{a\tilde
a}\psi_{a\tilde a}$ for any $\tilde a\leq a$. Finally observe that, 
by the defining equation of $\psi$ and by (\ref{A:1}) we have 
\begin{align*}
\psi_{a\tilde a}\,\psi_{\tilde a\hat a}\, \pi_{\hat a} (A)\,\Omega_{\hat a} 
    & =  \psi_{a\tilde a}\,\pi_{\tilde a } (\jmath_{a\tilde a}
  (A))\,\Omega_{\tilde a}
    =   \pi_{a } (\jmath_{a\tilde a}\,\jmath_{\tilde a\hat a}(A))
       \,\Omega_{a}\\
    & =   \pi_{a } (\jmath_{a\hat a}(A))
       \,\Omega_{a}
    =  \psi_{a\hat a}\,  \pi_{\hat a } (A)
       \,\Omega_{\hat a}
\end{align*}
%
for any $\hat a\leq \tilde a\leq a$ and $A\in \cA(\hat a)$,
and this implies that $\psi_{a\tilde a}\,\psi_{\tilde a\hat a} =
\psi_{a\hat a}$. So the pair $\{\pi,\psi\}$ is a net representation.\smallskip
  
In the present paper we are interested in unitary 
net representations. A net representation $\{\pi,\psi\}$ 
is said to be \emph{unitary} whenever 
$\psi_{a \tilde a}$ is a unitary operator for any $\tilde a\leq a$.
An interesting feature is that, since $K$ is pathwise connected, 
any unitary net representation is equivalent to a unitary net representation on a fixed Hilbert space. The argument is the same as that used to prove that any net bundle has a standard fibre 
(see \cite[Proposition 4.5]{RRV}).
Since the operators $\psi_{a \tilde a}$ are unitary, by a standard argument, one defines a unitary operator $V_a$ from a fixed Hilbert space 
$\cH$ to $\cH^\pi_a$ for any $a\in K$. Afterwards one defines  
\[
 \rho_a \doteq V^*_a\, \pi_a\, V_a\ , \ \  a\in K\ , \qquad 
 \phi_{a\tilde a}  \doteq V^*_a\, \psi_{a\tilde a}\, V_{\tilde a}\ , \ \  
 \tilde a\leq a\ .
\]
The pair $\{\rho,\psi\}$ is a unitary net
representation on the Hilbert space $\cH$;
the function $V: K\ni a\to V_a$
defines a unitary intertwiner from 
$\{\rho,\phi\}$ to $\{\pi,\psi\}$. From now on we will consider 
only unitary net representations in a fixed Hilbert space.\smallskip 

We denote by $\mathrm{Rep}^{net}(\cA)$ the set of unitary net representations 
of $\cA$ and by the same symbol the category 
having  unitary net representations 
of $\cA$ as objects and the corresponding intertwiners as 
arrows. We call this one  the category 
of \emph{unitary net representations} of $\cA$. If the target of an arrow 
$T$ is equal to the source of $S$, the composition 
$S\cdot T$ is defined by $(S\cdot T)_a = S_a\,T_a$ for any $a\in K$. 
The identity arrow 
$1_{\{\pi,\psi\}}$ is $(1_{\{\pi,\psi\}})_a=\mathbbm{1}_{\cH^\pi}$ for
any $a$. Furthermore, $\mathrm{Rep}^{net}(\cA)$ is a $\rC^*$--category. 
The adjoint ${}^*$ is defined as the identity on objects, 
while on arrows $T$ it is defined as 
$(T^*)_a=T^*_a$ for any $a$. Finally, 
given an arrow $T$, then 
$\norm{T}\doteq\sup_{a\in K}\norm{T_a}$  is a norm which makes 
$\mathrm{Rep}^{net}(\cA)$ a $\rC^*$--category. Note that by
(\ref{A:3}), since $K$ is pathwise connected,
$\norm{T_a}$ is constant.\smallskip

We now use the cohomology of posets 
to make explicit the topological content of 
unitary net representations. We give only a brief introduction 
of this topic  
and refer the reader to the  papers  
quoted at the beginning for details. Consider a Hilbert space $\cH$. 
A \emph{1--cocycle} $z$ of the poset $K$, 
with values in $\Bh$, is a function 
$z:\Si_1(K)\ni b\rightarrow z(b)\in\Bh$ 
of unitary operators of $\cH$ 
satisfying the equation 
\begin{equation}
\label{A:4}
z(\partial_0 c) \, z(\partial_2 c) =  z(\partial_1 c)\ , \qquad    
c\in\Si_2(K)\ . 
\end{equation}
The \emph{trivial} 1--cocycle $\imath$ is defined 
by $\imath(b)=\mathbbm{1}_{\cH}$ for any 1--simplex $b$. A 1--cocycle is a \emph{1--coboundary} if there is a function 
$v:\Sigma_0(K)\ni a\to v_a\in\Bh$ of unitary operators such that 
$z(b)= {v}^*_{\partial_0b}\,{v}_{\partial_1b}$ for any 1--simplex $b$. We
denote the set of 1--cocycles by $\rZ^1(K,\Bh)$. 
Given a pair $z,\tilde z$ of 1--cocycles an \emph{intertwiner} 
$t$ from $z$ to $\tilde z$ is a function 
$t:\Si_0(K)\ni a\to t_a\in\Bh $
satisfying 
\begin{equation}
\label{A:5}
t_{\partial_0b}\,  z(b) =  \tilde z(b)\,  t_{\partial_1b} \ , \qquad   
b\in\Si_1(K)\ .
\end{equation}
We denote the set of intertwiners from  $z$ to  $\tilde z$ by 
$(z,\tilde z)$.  The category of 1--cocycles  is the category 
whose objects are 1--cocycles and whose arrows are the corresponding 
set of intertwiners. We denote this category by the  
same symbol $\rZ^1(K,\Bh)$ as that used to denote the corresponding set 
of objects. This is  a $\rC^*-$category: composition of arrows and the
adjoint are defined in the same way as in $\mathrm{Rep}^{net}(\sA_K)$
(see \cite{Rob90,Ruz3} for details). 
Two 1--cocycles $z,\tilde z$ are  unitarily  \emph{equivalent} 
if there exists a unitary arrow $t\in(z,\tilde z)$.  Observe that 
any 1--coboundary is unitarily equivalent to the trivial 
1--cocycle $\imath$. \smallskip

We extend a 1--cocycle $z$
from 1--simplices to paths by setting 
\[
z(p)\doteq z(b_n)\cdots z(b_2)\, z(b_1)\ , \qquad 
p=b_n*\cdots*b_1\ .
\]
It is easily seen that $z(\overline{p})= z(p)^*$ for any path $p$, and
if $p$ and $q$ are homotopic  then 
$z(p)=z(q)$ (\emph{homotopic invariance}).
These properties imply that any 1--cocycle defines a unitary
representation, denoted by $z$, in $\cH$ of the fundamental group of the poset. Using this result the topological content of a unitary net
representations is easily analyzed. Indeed, 
given a unitary net representation $\{\pi,\psi\}$, define 
\begin{equation}
\label{A:6}
 \zeta^\pi(b)\doteq \psi^*_{|b|,\partial_0b}\,\psi_{|b|,\partial_1b}\ , \qquad 
     b\in\Si_1(K)\ .
\end{equation}
One can check that $\zeta^\pi$ is a 1--cocycle of $K$ with  
values in the group of unitary operators 
of $\Hil_\pi$ (\cite{RRV}). This 1--cocycle defines a representation of 
the fundamental group  of the poset. Thus, 
we say that a net representation $\{\pi,\phi\}$ 
is \emph{topologically trivial} whenever $\zeta^\pi$ 
is a 1--coboundary. Thus, 
if $K$ is simply connected then any unitary net representation 
is topologically 
trivial. 
\begin{lemma}
\label{A:7}
Assume that $\{\pi,\psi\}$ and $\{\rho,\phi\}$ are unitarily equivalent. 
Then the corresponding 1--cocycles $\zeta^\pi$ and $\zeta^\rho$ are
equivalent.
\end{lemma}
\begin{proof}
Let $W\in (\{\pi,\psi\},\{\rho,\phi\})$ be unitary. Then 
\begin{align*}
W_{\partial_0b} \, \zeta^\pi(b) & = 
W_{\partial_0b}\,\psi^*_{|b|,\partial_0b}\,\psi_{|b|,\partial_1b} 
= (\psi_{|b|,\partial_0b}\,W^*_{\partial_0b})^*\,\psi_{|b|,\partial_1b} \\
& = (W^*_{|b|}\,\phi_{|b|,\partial_0b})^*\,\psi_{|b|,\partial_1b} 
 = \phi_{|b|,\partial_0b}^*\, W_{|b|}\, \psi_{|b|,\partial_1b} \\
& = \zeta^{\rho}(b)\,W_{\partial_1b}\ .
\end{align*}
Hence $W\in(\zeta^\pi,\zeta^\rho)$ and this proves the assertion.
\end{proof}
Thus, equivalent unitary net representation have the same
topological content (the converse, in general, does not hold 
as we will see  at the end of this section). 
\begin{lemma}
\label{A:8}
Let $\{\pi,\psi\}$ be a topologically trivial 
unitary net representation. Then $\{\pi,\psi\}$ is  equivalent 
to a unitary net representation of the form $\{\rho,\mathbbm{1}\}$.
\end{lemma}
\begin{proof}  
Since $\zeta^\pi$ is a 1--coboundary there exists a family of unitary 
operators $W_a:\Hil\to \Hil$ such that 
$\zeta^\pi(b)= {W_{\partial_0b}}^*\, W_{\partial_1b}$. For
any 0--simplex $a$,  define  
$\rho_a(A) \doteq W_a\,\pi_a(A) \, {W_a}^*$, with $A\in\cA(a)$.
It is clear that $W_a\pi_a = \rho_a W_a$. Moreover, 
$W_a \,\psi_{a,\tilde a}  =  W^a \,\zeta^\pi(a;a,\tilde a)  =
W_{\tilde a}$, where $(a;a,\tilde a)$ is the 1--simplex whose support 
is $a$, and whose 0-- and 1--face are respectively 
$a$ and $\tilde a$. This completes the proof.
\end{proof}
A unitary net representation can be easily defined 
starting from a representation $\chi$ of the fundamental group 
of $K$ with values in the complex number $\mathbb{C}$. 
It is  shown in \cite{Ruz3} that  there is a 1--cocycle,
associated with $\chi$. We maintain the symbol $\chi$ to denote 
this 1--cocycle and define (see the previous proof for notation)
\begin{equation}
\label{A:9}
\psi^\chi_{\tilde aa}\doteq \chi(\tilde a;\tilde a,a)\ ,\qquad a\leq \tilde a\ .
\end{equation}
Consider now a topologically trivial 
net representation  $\{\pi,\mathbbm{1}\}$. Since 
$\psi^\chi$ takes values in the complex numbers, the pair 
$\{\pi,\psi^\chi\}$ is a unitary net representation 
(note that 
$\{\pi,\psi^\chi\}$  and $\{\pi,\mathbbm{1}\}$ are not equivalent
because  of Lemma \ref{A:7}). So,
if the fundamental group of the poset 
is Abelian, then there are  topologically non-trivial 
unitary net representations (clearly if the net is not trivial). 
For the non-Abelian 
case, topologically non-trivial examples, which are of interest for the theory 
of superselection sectors, will be given in Section \ref{F}.
Finally, note that the above example shows that 
there are non equivalent unitary net representations whose 
1--cocycles are equivalent. \bigskip

We conclude this section with some  observations. \smallskip  

$(1)$  A net  is nothing but a 
\emph{precosheaf}. By reverting the arrows,  the 
results of this section apply also to the duals, i.e. to presheaves, 
either of $\rC^*$--algebras or of groups. 
Briefly,  given a presheaf over a poset, 
if  we have a presheaf representation  on a Hilbert space 
whose {\em restriction} morphisms are implemented
by unitary operators satisfying relations corresponding to (\ref{A:2}), 
then the  presheaf representation carries a representation of the 
fundamental group of the poset. Proceeding as done 
for net representations, one gets 
a 1--cocycle of the dual poset $K^{\circ}$ (the poset having the same
elements as $K$ with opposite order relation). However, as shown 
in \cite{RRV}, the fundamental group of $K$ is isomorphic 
to that of $K^\circ$.\smallskip


$(2)$ The notion of representation of a net of $\rC^*$--algebras,
usually considered in the applications to quantum fields theory 
(see for instance \cite{HK,DHR,BF,GLRV,Ruz3,BR}), corresponds, in our framework, 
to a topologically trivial net representation. 
To see this note that, in the cited papers, 
by a representation of a net $\cA_K$ it is meant 
a function $\pi$ associating to any $a\in K$ 
a representation of $\cA(a)$ in a fixed Hilbert space $\Hil^\pi$, and
such that $\pi_{a}\,\jmath_{a\tilde a}=\pi_{\tilde a}$ for 
any $\tilde a \leq a$. An intertwiner
$S$ from a representation $\pi$ to a representation $\rho$, is a bounded
linear operator $S:\Hil^\pi\to \Hil^\rho$ such that 
$S\pi_a = \rho_aS$ for any $a\in K$. Now, it is clear 
that a representation is a unitary net representation of the
form $\{\pi,\mathbbm{1}\}$. Moreover, if  $T$ is an intertwiner 
from $\{\pi,\mathbbm{1}\}$ to  $\{\rho,\mathbbm{1}\}$,
then, according to (\ref{A:3}), we have 
$T_{a}= T_{\tilde a}$ for any $\tilde a\leq a$. Since $K$ 
is pathwise connected, we have $T_a= T_{\hat a}$ 
for any pair $a,\hat a$. So $T$ is constant. 
This shows that our definition is indeed a generalization, and 
in particular that the category of representations 
is equivalent to the full subcategory 
of $\mathrm{Rep}^{net}(\cA_K)$ 
whose net representations are topologically trivial.
We will denote this category by $\mathrm{Rep}^{net}_t(\cA_K)$. \smallskip

$(3)$ Carpi, Longo and Kawahigashi have considered 
representations of net over the covering spaces of $S^1$ \cite{CKL}.  
We think that unitary net representations are of the same 
nature of those considered by them. We prefer however  
not to explore this topic in the present paper.\smallskip


\section{Charged sectors induced by topology} 
\label{B}
In \cite{DHR}, Doplicher, Haag and Roberts were 
able to select a class of superselection sectors of the observable
net which manifest a covariant charge structure. 
These sectors, known as DHR--sectors, 
are representations of the observable net, in Minkowski spacetime, 
which are sharp excitations of the vacuum representation.  
This feature has been used to extend the notion of a DHR--sector 
to curved spacetimes \cite{GLRV}. In  4--dimensional globally 
hyperbolic spacetimes, DHR--sectors have 
a charge structure \cite{GLRV,Rob00,Ruz3},  which is 
generally covariant \cite{BR}. It is now clear 
that these sectors are not induced by the topology of spacetimes 
since they are associated with representations of the observable net
which are, in the terminology introduced in the present paper, 
topologically trivial.
In this section, by taking into account 
unitary net representations, we try to check whether they lead to 
genuine superselection structures.
We start by  discussing 
aspects of the causal structure of globally hyperbolic 
spacetimes, introduce the  
observable net, and the reference unitary net representation of the theory.
The definition of sharply localized unitary net 
representations concludes the section. For simplicity, 
from now on by a net representation we will always mean a unitary net 
representation.


\subsection{The observable net}
\label{Ba}

We start by discussing some aspects of the causal structure 
of globally hyperbolic spacetimes. The focus is on the set of 
diamonds, the class of regions that we will use as indices of the 
observable net. Standard properties 
of globally hyperbolic spacetimes can be found 
\cite{EH,BEE,One,Sen,Wal}. Some advanced aspects 
can be found in  \cite{BS05,BS06,MS06}.\bigskip 

Consider a 4--dimensional connected globally hyperbolic spacetime $M$. 
The \emph{causal disjointness} relation  is a symmetric 
binary relation $\perp$ defined on subsets of $M$ as follows: 
\begin{equation}
\label{Ba:1}
o\perp\tilde o \ \ \iff \ \ o\subseteq M\setminus J(o)\ ,
\end{equation}
where $J$ denotes the causal set of $o$. The causal complement 
of a set $o$ is the open set $o^\minperp\doteq M\setminus cl(J(o))$. An
open set $o$ is causally complete whenever $o=o^{\minperp\minperp}$. 
Now,  the observable net over the spacetime 
$M$ is a correspondence from open sets of the spacetime to 
the observables localized within these regions. 
In general not all the open sets are suited for this scope, since 
one needs a family of sets which fits very well 
both the topological and the causal properties of $M$. Moreover, 
additional conditions are imposed by the study of the observable nets 
derived from models of quantum fields \cite{Ve,Ruz2}. 
A family of sets that satisfies all these 
requirements is the set $K(M)$ of  \emph{diamonds}  of $M$ \cite{Ruz3}. 
A diamond of $M$ is a  subset 
$o$ of $M$ such that there is a spacelike Cauchy surface $\cC$, 
a chart $(U,\phi)$ of $\cC$, and an open ball $B$ of $\mathbb{R}^3$ 
such that 
\begin{equation}
\label{Ba:2}
 o = D(\phi^{-1}(B))\ , \ \ \ cl(B)\subset \phi(U)\subset\R^3\ ,
\end{equation}
where $D(\phi^{-1}(B))$ is the domain of dependence of $\phi^{-1}(B)$, 
and such that $cl(o)$ is \emph{compact}. 
We will say that $o$ is \emph{based} on $\cC$ and call 
$\phi^{-1}(B)$ the \emph{base} of $o$.
It turns out that, a diamond is an open, relatively compact, 
connected and simply connected subset of $M$. 
Any diamond $o$ is causally complete, and the causal complement  
$o^\minperp$ is  connected. The set 
of diamonds $K(M)$ of $M$ is a base for the topology of $M$.
Some technical properties of diamonds are 
shown in Appendix \ref{Z}. Notice that our present definition of diamonds differs, by the request of compactness of the closure, from the original one in \cite{Ruz3}. The results provided there and in \cite{BR} do not change after restriction to this smaller class.\smallskip 

We call a \emph{subspacetime} of $M$ 
any globally hyperbolic open connected subset of $M$. Diamonds 
and their causal complements are examples of subspacetimes of $M$.
Another example is the \emph{causal puncture} $x^{\minperp}$ in  
a point $x\in M$. This is nothing but the causal complement of 
the point $x$.
Now, it is an easy consequence \cite{BR} of
a powerful  result  on the deformation of Cauchy surfaces
\cite{BS06}, that 
\begin{equation}
\label{Ba:3}
 K(N) = \{ o\in K(M) \ : \  cl(o)\subset N\}\ ,
\end{equation}
for any subspacetime $N$ of $M$.\smallskip

We now move toward the definition of the observable net, and 
consider the poset formed by the set of diamonds 
of $M$ ordered under inclusion $\subseteq$.  
Some topological information of the spacetime can be deduced from 
the poset $K(M)$.  First of all, the poset $K(M)$ is 
pathwise connected since $M$ is connected. 
Secondly,  the first homotopy group of $K(M)$ is isomorphic 
to the first homotopy group of $M$. Furthermore,
recall that if a poset is upward directed, then  
it is simply connected (see Section \ref{A}). 
Thus $K(M)$ is not upward directed, 
when $M$ is not simply connected. The same happens when 
$M$ has compact Cauchy surfaces. \smallskip

The observable net in the Minkowski spacetime 
is defined according to the Haag-Kastler axioms \cite{HK}
(see also \cite{Ha}).
A generalization to a 4--dimensional globally hyperbolic spacetime $M$
has been provided in \cite{GLRV}. The \emph{observable net} 
$\sA_{K(M)}$ is defined as a   correspondence 
\begin{equation}
\label{Ba:4}
 o\mapsto \cA(o)\ ,
\end{equation}
associating with any diamond $o$ of $M$ a  unital 
$\rC^*$-algebra $\cA(o)$ representing  the algebra generated by 
\emph{all} the observables localized within $o$, and satisfying 
the \emph{isotony relation}
\begin{equation}
\label{Ba:5}
  o\subseteq \tilde o \ \ \Rightarrow \ \ \cA(o)\subseteq \cA(\tilde o)\ . 
\end{equation}
Isotony implies that the observable net
is a net of $\rC^*$--algebras over the poset $K(M)$.
Now, the Haag-Kastler axioms 
include the  Einstein's causality principle,
saying that observables localized in causally disjoint (spacelike separated)  
regions must commute. However, when the
indices of the observable net is a non-upward directed poset 
this principle cannot be fully implemented. Recall that  $K(M)$
fails to be upward directed 
when the spacetime  is not simply connected or when 
it has compact Cauchy surfaces. Following \cite{GLRV}, 
we restore this principle to the level of net representations 
of the observable net. A net representation $\{\pi,\psi\}$ 
of $\sA_{K(M)}$ is said to be \emph{causal}
whenever 
\begin{equation}
\label{Ba:6}
o\perp \tilde o \ \ \Rightarrow \ \ \pi_{o}(\cA(o))\subseteq 
\pi_{\tilde o}(\cA(\tilde o))', 
\end{equation}
where the prime stands for the commutant of the algebra.

\subsection{Sharply localized net representations: a selection criterion}
\label{Bb}

We start by introducing  the reference net representation. 
This net representation, 
that turns out to be a DHR--like representation because of the request of topological triviality, shall play for the theory 
the same r\^ole as  the vacuum one in  Minkowski spacetime. 
We conclude by giving the 
definition of net representations which are a sharp excitation
of the reference one. \smallskip


As a \emph{reference} net representation of 
the observable net   we consider a  faithful, \emph{causal 
and topologically trivial} net representation 
in an infinite separable complex Hilbert space $\cH_0$.
Thus, according to Lemma \ref{A:8}, we take
a net representation of $\sA_{K(M)}$
of the form  $\{\io,\mathbbm{1}\}$. Moreover, 
let $\sR_{K(M)}$ be  the net of von~Neumann algebras
\[
 o\mapsto \cR(o)\ , 
\]
where $\cR(o)\doteq \io_o(\cA(o))''$, that is, the observable net 
in the reference representation. Note that because of causality 
 if $o\perp \tilde o$ then $\cR(o)\subseteq \cR(\tilde o)'$. 
Then,  we require  that $\sR_{K(M)}$
satisfies the following properties.\bigskip 

\noindent \emph{Irreducibility} :  
$\mathbb{C}\,\mathbbm{1} =  \cap \{\cR(o)' \, |  \,  o\in K(M)\}$;\bigskip

\noindent \emph{Outer regularity} :  $\cR(o)= \cap \{\cR(\tilde o) \, |
\, cl(o)\subset \tilde o\}$;\bigskip

\noindent \emph{Borchers property} :  
given $o\in K(M)$ and a non-zero orthogonal 
projection $E\in\cR(o)$, for any $\tilde o$ with 
$cl(o) \subset  \tilde o$ there exists an isometry 
$V\in\cR(\tilde o)$ such that $VV^*= E$;\bigskip

\noindent \emph{Punctured Haag duality} :  
given a point $x\in M$ there holds 
\[
 \cR(o)
= \cap \big\{\cR(\tilde o)' \, : \,  \tilde o\in K(M), \ \tilde o \perp o, \ \ 
         cl(\tilde o)\perp \{x\}\big\},
\]
for any $o\in K(M)$ with $cl(o)\perp \{x\}$.\bigskip

Apart from the explicit request 
of topological triviality and outer regularity, the reference
representation  is defined in the same way as in  \cite{Ruz3}. 
Outer regularity, in particular, enters the theory only 
at one point, namely at the equivalence between 
sharply localized representations and net cohomology (see Appendix \ref{X}). 
We stress that physically meaningful examples of representations 
satisfying the defining properties of the reference representation
are the representations of a free scalar field which satisfy   
the microlocal spectrum condition \cite{Ve,Ruz2}, a generalization 
to globally hyperbolic spacetimes of 
the spectrum condition \cite{BFK,Rad}.\smallskip

As a consequence of the above assumptions (see \cite{Ruz3}) 
the net $\sR_{K(M)}$ satisfies 
Haag duality, i.e., 
$\cR(a) = \cap \big\{\cR(\tilde a)'\, :\, \tilde a \perp a\}$, for any 
$a\in K(M)$; and it is 
locally definite,  i.e., 
$\mathbb{C}\,\mathbbm{1} = 
  \cap\{ \cR(o) \ : \ x\in o\}$ for any $x\in M$.
Punctured Haag duality can be better understood by looking 
at the restriction of the theory to the causal punctures of the 
spacetime (see Section \ref{Ba}).
Let $\sR_{K(x^\minperp)}$ be the net obtained by restricting $\sR_{K(M)}$ 
to the set
of diamonds $K(x^{\minperp})$ with $x\in M$. 
Then $\sR_{K(x^{\minperp})}$ is an irreducible net 
satisfying Haag duality. We recall that the restriction 
to the causal punctures was the  key idea for the understanding
of the charge structure of DHR-sectors on globally hyperbolic spacetimes, 
mainly because the point 
$x$ plays for the set  $K(x^{\sst{\perp}})$ the same r\^ole 
as the spacelike infinity in  Minkowski spacetime. \smallskip

The purpose now is to generalize the criterion, used  
in \cite{GLRV,Ruz3} to select DHR-sectors, to net representations. 
DHR-sectors are topologically trivial net representations 
$\{\pi,\mathbbm{1}\}$  of $\Al_{K(M)}$ which are a sharp 
excitation of the reference, in symbols 
\begin{equation}
\label{Bb:0}
   \pi\restriction o^{\minperp} \cong \io\restriction o^{\minperp},  
\end{equation}
for any $o\in K(M)$. This means that for any $o$ there is 
a unitary operator $U^o:\cH^\pi\to \cH$ such that 
$U^o\,\pi_a = \io_a\,U^o$ for any $a\perp o$. 
This definition \emph{does not 
work}  if one takes into account  net representations 
which are not topologically trivial. In fact 
assume that $\{\pi,\psi\}$ is topologically non-trivial. 
If $\{\pi,\psi\}$ were equivalent to $\{\io,\mathbbm{1}\}$ on $o^{\minperp}$,
by Lemma \ref{A:7} the 1--cocycle $\zeta^\pi$ 
would be trivial on $o^\minperp$ and this leads to a contradiction. 
Indeed, let $\ell$ be a loop 
of $K(M)$ over a 0--simplex whose closure is contained in 
$o^\minperp$. By Corollary \ref{Z:6} $\ell$ is homotopic
to a loop $\ell'$ whose support has closure contained in $o^\minperp$. Then by 
homotopic invariance of 1--cocycles 
$\zeta^\pi(\ell)=\zeta^\pi(\ell')= 0$. Hence $\zeta^\pi$ should 
be trivial on $K(M)$.\smallskip


This  observation suggests how to modify (\ref{Bb:0}): 
we shall require that the above criterion is satisfied 
only in restriction to simply connected subspacetimes of $M$ 
(see Section \ref{Ba}). 
To be precise, we say that a causal net representation $\{\pi,\psi\}$ 
is a \emph{sharp excitation} of the reference one, if  
for any $o\in K(M)$ and 
for any \emph{simply connected} subspacetime $N$ of $M$, 
such that $cl(o)\subset N$, there holds 
\begin{equation}
\label{Bb:1}
 \{\pi,\psi\}\restriction o^{\minperp} \cap N  \ \cong \  
 \{\io,\mathbbm{1}\}\restriction o^{\minperp}\cap N\ .
\end{equation} 
This amounts to saying that 
there is a family 
$W^{N o}\doteq\{ W^{No}_a \,: \, cl(a)\subset N, \ a\perp o\}$ 
of unitary operators from $\cH^\pi$ to $ \cH_0$ such that 
\begin{enumerate}
\item  $W_a^{N o}\,\pi_a  = \io_a \,   W^{N o}_a$ ;
\item  $W_a^{N o}\,\psi_{a\tilde a}   = W_{\tilde a}^{N o}$\, for any  
  $\tilde a \subseteq a$ ;
\item $W^{N o}\, = {W^{N^{\prime} o}}$ for any simply connected 
subspacetime $N^\prime$ with $N\subseteq N^\prime$. 
\end{enumerate}
These three equations represent the \emph{selection criterion}. 
Observe that while equations 1 and 2 derive from (\ref{Bb:1})
and from  the definition of equivalent net representations, 
equation 3 does not. The latter equation is a compatibility 
request.\smallskip


Our next aim is to prove that this criterion is indeed a 
generalization of  (\ref{Bb:0}). Consider a causal representation 
of the form $\{\pi,\mathbbm{1}\}$ satisfying the above selection criterion. 
Given $o$ and $N$ as above, since $N$ is pathwise
connected then $W_a^{N o}$ is constant, i.e. independent of $a$ 
(see the first observation at the end of Section \ref{A}). 
So we can rewrite it as $W^{N o}$. By the third equation of the selection 
criterion we have  
\[
W^{N' o} = W^{N o}= W^{\tilde N o}\ , \qquad N\subseteq N'\cap \tilde N\ .
\]
This observation and, once again, pathwise connectedness of $M$
implies that $W^{N o}$ is independent of the region $N$. 
So we have obtained the DHR notion  of sharp excitation 
for topologically trivial representations. \smallskip
 
Denote the set of representations 
satisfying the selection criterion by $\rS\rC(\sA_{K(M)})$,
and consider the $\rC^*$--subcategory of $\mathrm{Rep}^{net}(\sA_{K(M)})$ 
whose set of objects is $\rS\rC(\sA_{K(M)})$. 
We denote this category by the same symbol 
$\rS\rC(\sA_{K(M)})$ used to denote the corresponding set of
objects. We denote  the full $\rC^*$--subcategory 
of $\rS\rC(\sA_{K(M)})$ whose objects are topologically trivial 
net representations by $\rS\rC_t(\sA_{K(M)})$. Because of  
the Borchers property, 
a routine calculation shows that 
these two  categories are  closed under direct sums and subobjects.  
Unitary equivalence classes of irreducible objects of 
$\rS\rC(\sA_{K(M)})$ are the \emph{superselection sectors} of the theory, 
and the analysis of their charge structure and topological 
content will be our scope from now on. 
Note that the superselection sectors of the subcategory $\rS\rC_t(\sA_{K(M)})$
are the DHR-sectors analyzed in \cite{Ruz3}.


\section{Net cohomology and the localization of the fundamental group}
\label{C}
The category of the net representations satisfying the
selection criterion admits an equivalent description in terms of the net 
cohomology of the poset $K(M)$ with values in the 
observable net. 
In the present section we prove a property of net 
cohomology which is at the base of this equivalence,
the localization of the fundamental group, i.e.,  
the representation of the first homotopy 
group defined by a 1--cocycle is localized.
As we shall see in the following, this property  is the key 
for understanding both charge structure and topological content 
of superselection sectors. \bigskip

Consider the observable net in the reference 
representation $\sR_{K(M)}$, and 
denote the category of 1--cocycles of the
poset $K(M)$ with values in the net $\sR_{K(M)}$  by
$\rZ^1(\sR_{K(M)})$. 
This is 
the $\rC^*$--subcategory of $\rZ^1(K(M),\mathfrak{B}(\cH_0))$ 
whose objects $z$ and whose arrows $t\in (z,\hat z)$ 
satisfy the \emph{locality condition}, i.e., 
\begin{equation}
\label{Ca:1}
 z(b)\in \cR(|b|)\ , \qquad b\in\Si_1(K(M))\ ,
\end{equation}
and
\begin{equation}
\label{Ca:2}
 t_a\in \cR(a)\ , \qquad a\in\Si_0(K(M))\ .
\end{equation}
Now, as any 1--cocycle
defines a representation of the fundamental group 
of the poset $K(M)$, we denote the full $\rC^*$--subcategory of 
$\rZ^1(\sR_{K(M)})$ whose objects are trivial representations 
of the fundamental group by $\rZ^1_t(\sR_{K(M)})$. Sometimes
we shall refer to the elements of  $\rZ^1_t(\sR_{K(M)})$
as topologically trivial 1--cocycles. Note that topologically trivial 
1--cocycles are nothing but that  1--coboundaries 
in  $\rZ^1(K(M),\mathfrak{B}(\cH_0))$. \smallskip

Before diving further into the deep sea of net cohomology, some words of explanation are in order. As pointed out in Section \ref{Ba},
the set of indices of the observable 
net in a globally hyperbolic spacetime is  non-directed under inclusion
when the spacetime either is multiply connected or has compact 
Cauchy surfaces. In such  situations it is not possible to define 
the $\rC^*$--algebra of all local observables, 
i.e., the $\rC^*$-inductive limit. This does not happen in  Minkowski 
spacetime where there is a canonical choice of the set of indices,
the set of double cones, which is directed under inclusion. 
This fact  reflects in the way how DHR-sectors have been analyzed
in these two situations. A key step of DHR analysis 
was the understanding that  all 
the information of sectors is encoded  in 
a unique Hilbert space: that one related to the vacuum. 
The category of DHR-sectors 
in Minkowski spacetime turns out to be equivalent to the category of 
localized and transportable endomorphisms 
of the \emph{algebra of all local observables} defined   
in the vacuum representation \cite{DHR}.
In globally hyperbolic spacetimes
DHR-sectors are still  encoded in the vacuum Hilbert space (the reference) 
but in a different form. There is a notion 
of  localized and transportable endomorphisms of the \emph{observable net} 
defined in the vacumm representation, but the corresponding 
category is not equivalent to the category of DHR-sectors anymore when 
the set of indices is non-directed under inclusion \cite{GLRV}.
The functor from the latter to the former category is not full, and it 
is not clear how to define the functor 
in the opposite direction. However, 
as pointed out by Roberts, in Minkowski 
spacetime DHR-sectors can be equivalently 
described in terms of the operators that in DHR analysis play the r\^ole 
of charge transporters: 1--cocycles of 
double cones taking values in the observable 
net defined in the vacuum representation \cite{Rob76,Rob90}. 
As shown in \cite{GLRV}, this equivalence  maintains in arbitrary 
globally hyperbolic spacetimes:
the category $\rS\rC_t(\sA_{K(M)})$ is equivalent to the category
$\rZ^1_t(\sR_{K(M)})$ (see also \cite{Ruz3}).  
Analyzing  the category $\rZ^1_t(\sR_{K(M)})$,
the charge structure and the general covariance 
of DHR-sectors have been understood \cite{GLRV,Rob00,Ruz3,BR}.\smallskip
%

Our first step for understanding  sectors introduced in the present 
paper according to the selection criterion (\ref{Bb:1})
consists in proving that the category $\rS\rC(\sA_{K(M)})$ is equivalent 
to  $\rZ^1(\sR_{K(M)})$.  However, in this case,  we have to pay attention, 
since non trivial topological 
objects are involved. To begin with, we  show the property of 
net cohomology which underlies this equivalence.\smallskip


Let  $\pi_1(K(M),a)$ be the 
first homotopy group 
of the poset $K(M)$  based on the 0--simplex $a$. 
Given a 1--cocycle $z$ of $\rZ^1(\sR_{K(M)})$,  
we call the  von Neumann algebra defined by 
\begin{equation}
\label{Ca:3}
 \cR^z(M,a) \doteq  \{ z(\ell) \, :  \, \ell\in\Loops{K(M)}(a)\}'',
\end{equation}
the \emph{group algebra} associated with $z$, where the double prime 
stands for the bicommutant. The next theorem, 
to which we shall refer 
as the \emph{localization of the fundamental group}, 
asserts that this von Neumann algebra is localized.
\begin{theorem}
\label{Ca:4}
Given  $z\in \rZ^1(\sR_{K(M)})$, the following assertions hold :
\begin{itemize}
\item[(i)] $z(\ell)\in\cR(a)$,  $\forall\ell\in\Loops{K(M)}(a)$;
\item[(ii)] $\cR^z(M,a)\subseteq \cR(a)\ $. 
\end{itemize}
\end{theorem}
\begin{proof}
Note that if we prove that $z(\ell)\in \cR(o)'$ for any
diamond $o$ such that $o\perp a$, then Haag duality 
implies that $z(\ell)\in\cR(a)$. To this end, observe that 
if $o\perp a$ then, by Corollary \ref{Z:6}, the loop $\ell$
is homotopic to a loop $\ell_1$ whose 
support\footnote{The support 
of a path is the union of the supports of the 1--simplices 
that form the path.} is contained 
in the causal complement of $o$. By homotopic invariance 
of 1--cocycles we have  
$z(\ell)=z(\ell_1)\in \cR(o)'$ completing the proof.
\end{proof}
In \cite{GLRV} it was proved that any 1--cocycle $z$ 
of $\rZ^1_t(\sR_{K(M)})$ 
satisfies the following localization properties:
\emph{First}, given a path $p$, then 
\begin{equation}
\label{Ca:5}
z(p)\in\cR(o)',\qquad  o\perp \partial p\, ,
\end{equation}
where $\partial p$ denotes the boundary of the path, i.e.,  $\partial p=\{\partial_0p,\partial_1p\}$;
\emph{secondly}, let $p,q$ be paths with $\partial_0p=\partial_0q$, 
and let $o$ be a diamond  such that $\partial_1p,\partial_1q \perp o$, then 
\begin{equation}
\label{Ca:6}
z(p)\, A\, z(p)^* = z(q)\, A\,z(q)^*,\qquad A\in\cR(o)\ .
\end{equation}
Thanks to the localization of the fundamental group 
we now are able to prove that these properties hold in full generality. 
\begin{corollary}
\label{Ca:7}
Any 1--cocycle of $\rZ^1(\sR_{K(M)})$ satisfies 
(\ref{Ca:5}) and (\ref{Ca:6}).
\end{corollary}
\begin{proof}
Let us prove (\ref{Ca:5}). As the causal complement of $o$ is 
pathwise connected, there 
is a path $q$ with $\partial q = \partial p$ and $|q|\perp o$. 
Observe that $\overline{q}*p$ is a loop whose endpoint, say $a$, 
is causally disjoint from $o$. Since 
$z(\overline{q}*p)\in\cR(a)$ because of 
Theorem \ref{Ca:4}, we have 
$z(p)\, A$ $= z(q)\, z(\overline{q}*p)\, A$ 
$=z(q)\, A \, z(\overline{q}*p)$ 
$= A\, z(q)\, z(\overline{q}*p)$ 
$= A\, z(p)$, for any $A\in\cR(o)$ and this proves relation (\ref{Ca:5}).

Let $p,q$ and $o$ be as in  (\ref{Ca:6}). As 
$\overline{q}*p$ 
satisfies, with respect to $o$,  the hypotheses  of 
(\ref{Ca:5}), we have 
$z(p)\,A \, z(p)^*$
$=z(q)\, z(\overline{q}*p)\, A 
                            \, z(\overline{q}*p)^* \, z(q)^*=$
$z(q)\, A\, z(q)^*$, for any $A\in\cR(o)$.
\end{proof}
Now, the  first application of this corollary is the crucial equivalence
between sharply localized net representations and 
net cohomology.
\begin{theorem}
\label{Ca:8}
$\rS\rC(\sA_{K(M)})$ and $\rZ^1(\sR_{K(M)})$ are equivalent categories.
\end{theorem}
We prefer to postpone the, rather technical, proof of 
this equivalence in Appendix. We only point out 
that the functors that define 
the equivalence are an extension of the functors that define the equivalence 
between $\rS\rC_t(\sA_{K(M)})$ and $\rZ^1_t(\sR_{K(M)})$. 
On these grounds the \emph{superselection sectors} are described 
by the unitary equivalence classes of the irreducible objects 
of the category $\rZ^1(\sR_{K(M)})$. 
So from now on, the analysis of superselection sectors
will be carried out on $\rZ^1(\sR_{K(M)})$. \smallskip

Corollary \ref{Ca:7} applies also to the analysis of the charge structure of superselection sectors which is the subject of the next section.


\section{Charge Structure}
\label{D}

The purpose of the present section is to show that the superselection 
sectors previously introduced manifest a charge structure. 
As observed in the previous section, this analysis 
will be performed on the  category $\rZ^1(\sR_{K(M)})$.
At this point it is worth recalling that the charge structure  
of topologically trivial cocycles has been completely understood: 
the $\rC^*$--category  $\rZ^1_t(\sR_{K(M)})$ 
has a tensor product, a permutation symmetry and a conjugation. 
This amounts to saying that the quantum numbers, i.e., 
the labels of sectors, have a composition law, a particle-antiparticle 
symmetry and an additional number saying that a  sector 
has  either the para-Bose or the para-Fermi statistics. \smallskip 

The study of the category $\rZ^1_t(\sR_{K(M)})$  resembles a standard 
argument of differential geometry. 
One first restricts the attention to the causal punctures of the spacetimes,
namely  to the categories $\rZ^1_t(\sR_{K(x^{\sst{\perp}})})$.  
The advantage is that in these regions 
the point $x$ has properties similar to the spacelike infinite
in the Minkowski space. So one can prove 
the existence of a tensor product, permutation symmetry, left inverses 
and conjugated object in $\rZ^1_t(\sR_{K(x^{\sst{\perp}})})$ for any point $x$.
Then, one observes that for 1--cocycle $z$ of $\rZ^1_t(\sR_{K(M)})$
the local definitions, i.e. on the causal
punctures,  of tensor product, permutation symmetry and conjugated 
objects can be glued together to form the corresponding 
global notions. \smallskip

It is now important to note that most of the
constructions made for  $\rZ^1_t(\sR_{K(M)})$, 
the tensor product, the permutation symmetry and 
the conjugation do not involve the topological triviality of 
1--cocycles directly but rather  relations 
(\ref{Ca:5}) and (\ref{Ca:6}). Thus, by Corollary \ref{Ca:7},  
these constructions can be straightforwardly applied to  
$\rZ^1(\sR_{K(M)})$.
Only one point of that analysis  cannot be extended to the general case: 
the proof of the existence of left inverses (and consequently the definition of statistics) because it relies 
on the fullness of the restriction functor from $\rZ^1_t(\sR_{K(M)})$ to 
$\rZ^1_t(\sR_{K(x^{\sst{\perp}})})$, a  property that does not hold for 
general 1--cocycles. However,  although we will not prove the existence 
of left inverses for all the objects of $\rZ^1(\sR_{K(M)})$
we will be able to define objects with finite statistics via a 
detour.\smallskip

We assume that the reader is familiar with 
symmetric tensor $\rC^*$--categories and related notions. 
Two references for this topic are \cite{DR89,LR}.
Other references whose focus is on the theory 
of superselection sectors are  \cite{Rob90,BW,Rob00,HM}.


\subsection{DHR-like endomorphisms}
\label{Da}

Any 1--cocycle defines a class of endomorphisms that are localized 
and transportable in the same sense of those used in DHR analysis, but 
live on a presheaf associated with the observable net.  
Although these endomorphisms do not contain all the information about 
superselection sectors (see Remark \ref{Da:6a}), they enter  the definitions 
of tensor product, permutation symmetry and conjugation.\smallskip

Given a diamond $o$, the algebra of its \emph{causal complement} 
is the $\rC^*$-algebra  $\cR^\minperp(o)$ 
generated by all the algebras $\cR(a)$ with $a\perp o$. 
The \textit{presheaf} $\sR^\perp_{K(M)}$ 
associated with the observable net  is  the correspondence 
$o\to  \cR^\minperp(o)$.\\
\indent Consider a 1--cocycle $z$ of $\rZ^1(\sR_{K(M)})$. Fix a 0--simplex $o$,
and let $a$ be a 0--simplex such that $a\perp o$. Define 
\begin{equation}
\label{Da:2}
y^z_a(o)(A) \doteq z(p)\, A\, z(p)^*\ , \qquad A\in\cR^\minperp(a)\ ,
\end{equation}
where $p$ is path with $\partial_1p\subseteq a$ and 
$\partial_0p=o$. By (\ref{Ca:5}) and (\ref{Ca:6})
this definition does not depend on the path chosen $p$  
and on the choice of the starting point $\partial_1p$.
Therefore
\begin{equation}
\label{Da:3}
y^z_{\tilde a}(o)\upharpoonright \cR^\minperp(a)  = y^z_{a}(o)\ , 
\qquad \tilde a \subseteq a \ .
\end{equation}
\indent Fix a point $x$ of the spacetime $M$. Since $K$ is a base 
for the topology of $M$, the collection of $0$--simplices 
$K(x)\doteq \{\tilde a : x\in \tilde a\}$ is downward directed.
The \textit{stalk} in a point $x$ can be seen either as the 
$\rC^*$-inductive limit of the system 
$\cR^\minperp(o)$ with $o\in K(x)$ or as the 
$\mathrm{C}^*-$algebra generated 
by  the algebras $\cR(o)$ for any $o$ in 
$K(x^\minperp)$. \\
\indent Then, by property (\ref{Da:3}), the collection
\begin{equation}
\label{Da:4}
y^z_x(o)\doteq\{y^z_a(o) \,|\,  a\in K(x)\}\ , \qquad o\in K(x^\minperp) \ , 
\end{equation}
is extendible to a morphism of the stalk $\cR^\minperp(x)$.  
%
\begin{lemma}
\label{Da:lemma} 
On the premises outlined before, we have that 
$y^z_x(o)$ is an endomorphism of $\cR^\minperp(x)$
satisfying the following properties:
\begin{itemize}
\item[$(i)$]  $y^z_x(o) \restriction \cR(\tilde o) =
id_{\cR(\tilde o)}$ 
    for any $\tilde o\in K(x^\minperp)$ with $\tilde o\perp o\ $;
\item[$(ii)$] 
$z(p)\,  y^z_x(\partial_1p) =  y^z_x(\partial_0p)\,
    z(p)$ for any path  $p$ in $K(x^\minperp)$;
\item[$(iii)$]  
$t_o \,  y^z(o) = y^{z_1}(o)\, t_o$, with $t\in(z,z_1)$;
\item[$(iv)$]  $y^z(o)(\cR(\tilde o))\subseteq \cR(\tilde o)$ 
for any $\tilde o\in K(x^\minperp)$ with 
$o\subseteq \tilde o\ $.
\end{itemize}
\end{lemma}
%
The proof of these properties is the same as the proof of 
\cite[Lemma 4.5]{Ruz3}. We only observe that 
the first three properties are a consequence of the
localization of the fundamental group (Properties (\ref{Ca:5}) 
and (\ref{Ca:6})).  
Property $(iv)$ derives from property $(i)$ and 
from punctured Haag duality, because  the restriction 
of $\sR_{K(M)}$ to $K(x^\minperp)$ satisfies Haag duality 
(see Section \ref{Bb}).

\begin{remark}
\label{Da:6a}
The family $\{y^z_x(o) \,|\, o\in K(x^\minperp)\}$ of endomorphisms 
of $\cR^\minperp(x)$ are \emph{localized} $(i)$ and \emph{transportable} 
$(ii)$ in the same sense of DHR analysis. So, using the interpretation 
given in DHR analysis we may think of $y^z_x(o)$ as a charge localized 
within the diamond $o$ and the 1--cocycle 
$z$ as the transporter of these charges. As said at the beginning of this 
section, these endomorphisms do not contain all the information of 
superselection sectors. Among the various difficulties, the simplest one is
to observe that if $z$ is an irreducible object but carries a non-trivial 
representation of the fundamental group then, by property 
$(ii)$ of Lemma \ref{Da:lemma}, 
the corresponding endomorphism is not irreducible.  
\end{remark}

We finally point out  two useful relations.
The first one,  an obvious consequence 
of (\ref{Da:3}),  says that 
\begin{equation}
\label{Da:7}
 y^z_x(o) \restriction \cR^\minperp(a) =  y^z_{\tilde x}(o)  
\restriction \cR^\minperp(a)\ ,  
\end{equation}
where $x,\tilde x\in a$ and $o\in 
K(x^\minperp)\cap K(\tilde x^\minperp)$.
This, in turn, implies that
\begin{equation}
\label{Da:8}
 y^z_x(o)(z(p)) =  y^z_{\tilde x}(o)(z(p))\ ,
\end{equation}
for any pair $x,\tilde x$ of points and any path $p$ such that 
$|p|\subseteq K(x^\minperp)\cap K(\tilde x^\minperp)$.
The proof of these two properties is given in 
\cite{Ruz3} where they are called gluing conditions, because they allow to extend cocycles and arrows defined on causal punctures over all $K(M)$.


\subsection{Tensor structure}
\label{Db}
Thanks to the localization of the fundamental group, 
we shall define the tensor product and the 
permutation symmetry in $\rZ^1(\sR_{K(M)})$ by the same formulas 
as those used to define the corresponding notions in 
$\rZ^1_t(\sR_{K(M)})$ \cite{Ruz3}. In that paper 
these formulas are first defined on the causal punctures 
and after extended globally by the gluing conditions. 
For brevity we shall give directly the global definitions.
Clearly, most of the proofs are omitted, with some exceptions
because they need modifications from 
the original ones.\smallskip

We start by introducing a preliminary definition.
Given $z,z_1\in \rZ^1(\sR_{K(M)})$ and $t\in(z,z_1),s\in(z_2,z_3)$,  define 
\begin{equation}
\label{Db:1}
\begin{array}{rcll}
  z(p)\times_x z_1(q) &\doteq& z(p)\, y^z_x(\partial_1p)(z_1(q))\ , & 
   p,q \mbox{ paths in } K(x^\minperp)\ ,\\[5pt]
  t_a\times_x s_{\tilde a} &\doteq&  t_a \,
  y^z_x(a)(s_{\tilde a})\ , & a\in  \Si_0(K(x^\minperp))\ .
\end{array}
\end{equation}
As a consequence of properties $(ii)$ and $(iii)$ 
of localized transportable endomorphisms 
(see Lemma \ref{Da:lemma}) 
we have 
\begin{equation}
 \label{Db:2a}
z(p*\hat p)\!\times_x \!z_1(q*\hat q)  =  
       {z(p)\!\times_x \!z_1(q)} \ \, {z(\hat p)\!\times_x\!z_1(\hat
       q)}\ ,
\end{equation}
and 
\begin{equation}
\label{Db:2b}
t_{\partial_0p}\!\times_x \!s_{\partial_0q} \ 
       z(p)\!\times_x \!z_1(q)  =
 {z_2(p)\!\times_x \!z_3(q)} \  {t_{\partial_1p}\!\times_x\!
       s_{\partial_1q}}\ ,
\end{equation}
%
(cfr.  \cite[Lemma 4.6]{Ruz3}). Furthermore we have the following
\begin{lemma}
\label{Db:3}
Given a pair of paths  $p,q$  of  $K(x^\minperp)$. Then 
\[
z(p) \times_x z_1(q) = z_1(q) \times_x  z(p)\ , 
\]
whenever  $\partial_ip\perp\partial_iq$  for  $i=0,1$. 
\end{lemma}
\begin{proof}
There are in $K(x^\minperp)$ 
two paths $p_1 =b_{j_n}*\cdots *b_{j_1}$ and 
$q_1 =b_{k_n}*\cdots *b_{k_1}$  such that  
$|b_{j_i}|\perp|b_{k_i}|$ for $i=1,\ldots, n$
and  $\partial p_1=\partial p$, $\partial q_1=\partial q$,
see \cite[Section 3.2.1]{Ruz3}.
For these paths we have 
$z(p_1) \times_x  z_1(q_1) =  z_1(q_1) \times_x  z_1(p_1)$
(cfr. \cite[Lemma 4.8]{Ruz3}). 
Using this and (\ref{Db:2a}) we have 
\begin{align*}
z(p) \times_x  z_1(q) &  =   z(p_1) \times_x  z_1(q_1) \   
    z(\overline{p_1}*p) \times_x  z_1(\overline{q_1}*q)  \\
      &  =   z_1(q_1)\times_x z(p_1) \  
             z_1(\overline{q_1}*q) \times_x z(\overline{p_1}*p) \\
 &  =   z_1(q)\times z(p)\ ,
\end{align*}
where we have used the fact that 
\[
    z(\overline{p_1}*p) \times_x  z_1(\overline{q_1}*q)   = 
    z(\overline{p_1}*p)\,   z_1(\overline{q_1}*q)  
    =    z_1(\overline{q_1}*q) \,  z(\overline{p_1}*p)  =  
                z_1(\overline{q_1}*q) \times_x z(\overline{p_1}*p) \ , 
\]
because $z(\overline{p_1}*p) \in  \cR(\partial_1p)$, 
$z_1(\overline{q_1}*q) \in \cR(\partial_1q)$,
$\partial_1p\perp\partial_1q$, and property $(i)$ of 
localized and transportable endomorphisms of stalks.
\end{proof}
The tensor product 
is a particular case of the expressions (\ref{Db:1}).
Given $z,z_1\in \rZ^1(\sR_{K(M)})$ and $t,s$ arrows of 
$\rZ^1(\sR_{K(M)})$ define 
\begin{equation}
\label{Db:4}
\begin{array} {rcll}
  (z\otimes z_1)(b) & \doteq &  z(b)\times_x z_1(b)\ , &
	b\in\Si_1(K(M))\ , \\[3pt]
  (t\otimes s)_a  & \doteq &  t_a \times_{\tilde x} s_a\ ,  &  a\in\Si_0(K(M))\ ,
\end{array}
\end{equation}
where $x$ and $\tilde x$ are  points of $M$ such that 
$x\perp cl(|b|)$  and  $\tilde x\perp cl(a)$.
One first observes that these definitions 
behave as a tensor product when restricted to the causal puncture 
$K(x^\minperp)$ of $M$ in $x$. Afterwards one observes that 
by (\ref{Da:8}) the definitions are independent of the choice of the 
point (\cite[Proposition 4.7 and Lemma 4.17]{Ruz3}).\smallskip 

The following lemma characterizes the morphism of stalks 
associated with the tensor product of two 1--cocycles.
\begin{lemma}
\label{Db:5}
Let $z,\tilde z\in \rZ^1(\sR_{K(M)})$. 
Then $y^{z\otimes z_1}_x(o) = y^{z}_x(o) y_x^{z_1}(o)$, for
any $o\in K(x^\minperp)$.
\end{lemma}
\begin{proof}
Let $a$ be a 0--simplex in $K(x)$ (see Section \ref{Db}) such that 
$cl(a)\perp o$. Let $p$ be a path in $K(x^\minperp)$ 
such that $cl(\partial_1p)\subset a$ 
and  $\partial_0p=o$. Take 
$A\in\cR^\minperp(a)$. According to the definition (\ref{Da:4}) we have 
\begin{align*}
y^{z\otimes  z_1}_x(o)(A)&  = 
y^{z\otimes z_1}_a(o)(A) \\
& =(z\otimes z_1) (p)\, A\, (z\otimes z_1)(p)^* \\
& =
 (z(p)\times _x z_1(p)\big)\,A\,  
      \big(z(p)\times _x  z_1(p)\big)^* \\
&   = \big(z(p)\,y^z_x(\partial_1p)(z_1(p))\big) \,  A\,
       \big(z(p)\,y^z_x(\partial_1p)(z_1(p))\big)^*\\
 &  = z(p)\, z(q)\,  z_1(p)\,  z(q)^* \,  A\, 
       \big(z(p)\, z(q)\,  z_1(p)\,  z(q)^*\big)^*\ ,
\end{align*}
where $q$ is a path in $K(x^\minperp)$ such that 
$\partial_0q=\partial_1p$ and $cl(\partial_1q)\subset a$ 
and  $\partial_1q\perp\partial_1p$.\footnote{Note that such a path exists
since $cl(\partial_1p)\subset a$, by Lemma \ref{Z:4}, there are two diamonds 
$o_1$ and $o_2$ such that $cl(o_2)\subset a$ and 
$cl(o_1)\perp cl(\partial_1p)$. So we can take   
$q$ to be the 1--simplex whose support is $o_2$, the 1--face 
is $o_2$, and the 0--face is $\partial_1p$.}
Applying (\ref{Ca:5}) we have 
$z(q)^* \,  A\, z(q)=A$; thus 
\begin{align*}
y^{z\otimes z_1}_x(o)(A) &  = 
z(p)\, z(q)\,  z_1(p)\, A
      (z(p)\, z(q)\, z_1(p))^*\\
  &  = z(p*q)\,   z_1(p)  \, A\, 
       (z(p*q)\,  z_1(p))^*\\
  &  = z(p*q)  \, y^{z_1}_x(o)(A)\,  
       z(p*q)^*\\
  &  = y^z_x(o)\big(y^{z_1}_x(o)(A)\big) \ ,
 \end{align*}
where we have used the relation
$y^{z_1}_x(o)(A)\in\cR^\minperp(a)$ which is a consequence
of property $(iv)$ of localized and transportable endomorphisms 
of stalks. 
\end{proof}
The permutation symmetry $\eps$ is defined, for any pair 
$z,z_1$ in $\rZ^1(\sR_{K(M)})$, by  
\begin{equation}
\label{Db:6}
\eps(z,z_1)_a \doteq z_1(q)^*  \times_x z(p)^* \  z(p)  \times_x  z_1(q)\ ,
\qquad a\in\Si_0(K(M))\ ,  
\end{equation}
where $x$ is any  point of $M$ with $x\perp cl(a)$, and 
$p,q$ are two paths  of $K(x^\minperp)$ with  
$\partial_0p\perp\partial_0 q$ and   
$\partial_1p=\partial_1 q=a$. Once again one restricts the attention 
to the causal punctures. First one observes that 
the above definition does not depend on the choice of the paths 
$p$ and $q$, and shows 
$\eps$ is indeed a symmetry in restriction to the causal punctures. 
Afterwards, one checks that the above definition does not depend on the choice 
of the point $x$ (see \cite{Ruz3}). A useful relation 
for analyzing the topological content of 1--cocycles 
is provided in the following lemma. 
\begin{lemma}
\label{Db:7}
Given $o$ and $x\in M$ with $cl(o)\perp x$, then 
\[ 
\eps(z,z_1)_o \  z(\ell)\times_x z_1(\ell') =  
z_1(\ell') \times_x z(\ell) \ \eps(z,z_1)_o\ ,
\]
where $\ell,\ell' \in\Loops{K(x^\minperp)}(o)$.
\end{lemma}
\begin{proof}
Consider a point $x$ and two paths $p$ and $q$ as  
in the definition of $\eps$. By 
(\ref{Db:2a}) we have 
\begin{align*}
\eps(z,z_1)_o \    z(\ell)\times_x z_1(\ell') 
         & =  {z_1(q)^*  \!\times_x\! z(p)^*}  \   {z(p)\!\times_x\! z_1(q)}\ 
              {z(\ell)\!\times_x\! z_1(\ell')} \\ 
            & = {z_1(q)^*\!\times_x\! z(p)^*} \ { z(p*\ell)\! \times_x\!
              z_1(q*\ell')}\\
            & = {z_1(\ell')z_1(q*\ell')^*\!\times_x\!
                z(\ell) z(p*\ell)^*} \  z(p*\ell)  \times_x
              z_1(q*\ell')\\
             & = 
               {z_1(\ell')\!\times_x\! z(\ell)} \ 
                   {z_1(q*\ell')^*\!\times_x\!
                    z(p*\ell)^*} \  {z(p*\ell)\!\times_x\!z_1(q*\ell')}\\
             & = z_1(\ell')\times_x z(\ell) \  \eps(z,z_1)_o
\end{align*}
because $p*\ell$ and $q*\ell'$ are paths satisfying the definition 
of $\eps$.
\end{proof}
Note that if we take in this lemma $\ell'$ as the trivial loop, 
i.e., $\ell'=\si_0o$ then
$\eps(z,z_1)_o \,  z(\ell) =  
 y^{z_1}_x(o)(z(\ell)) \, \eps(z,z_1)_o$. Since the unitaries 
$z(\ell)$ generate the algebra $\cR^z(M,o)$ and since 
$y^{z_1}_x(o)$ is normal on this algebra  we have 
\begin{equation}
\label{Db:8}
A = \eps(z_1,z)_o\, y^{z_1}_x(o)(A) \, \eps(z,z_1)_o\ , \qquad 
A\in\cR^z(M,o)\ ,
\end{equation}
with $o\in K(x^\minperp)$.

\subsection{Statistics and Conjugation} 
\label{Dc}
Our purpose now is to identify the objects of 
$\rZ^1(\sR_{K(M)})$ having conjugates. The first 
step will be to understand what are the objects with 
finite statistics. To reach this goal  we shall not follow 
the traditional way, rather we shall identify 
a $\rC^*$--subcategory $\widetilde{\rZ}^1(\sR_{K(M)})$ 
closed under tensor product, direct sums, subobjects 
and having left inverses, and containing all  the simple objects 
of $\rZ^1(\sR_{K(M)})$. Within this category we shall define 
the objects with finite statistics in a same way as in DHR analysis. 
Afterwards, we prove that any object with finite statistics 
has conjugates.\smallskip

We recall that a left inverse $\phi$  of an object $z$ of a tensor 
$\rC^*$--category  is a family of linear mappings
$\phi_{z_1,z_2}:(z\otimes z_1,z\otimes
z_2)\to (z_1,z_2)$, for pair any $z_1,z_2$ of objects, satisfying 
the following relations: given $X\in(z\otimes z_1,z\otimes z_2)$, then
\begin{align*} 
\phi_{z_1\otimes \tilde z,z_2\otimes \tilde z}(X\otimes 1_{\tilde z})
  & =  \phi_{z_1,z_2}(X)\otimes 1_{\tilde z}\ ,\\
 \phi_{z^{\prime},z^{\prime\prime}}(1_z\otimes S\cdot X\cdot 1_z\otimes R) & = 
  S\cdot \phi_{z_1,z_2}(X) \cdot R \ , 
   \qquad  S\in (z_2,z^{\prime\prime}) \ ,  R\in (z_1,z^{\prime}) \ .
\end{align*}
A left inverse of $z$ is said to be \emph{positive} whenever, 
for any object $\tilde z$,
$\phi_{\tilde z,\tilde z}$ sends positive elements 
of $(z\otimes \tilde z,z\otimes \tilde z)$ in to positive elements 
of $(\tilde z,\tilde z)$; 
\emph{normalized} whenever $\phi_{\io,\io}(1_z) = 1_{\io}$.
A positive left inverse $\phi$ of $z$ is said to be 
\emph{faithful} whenever, for any object $\tilde z$, 
$\phi_{\tilde z,\tilde z}(X)\ne 0$ for any positive 
and non-zero element $X$ of $(z\otimes \tilde z,z\otimes \tilde z)$.\smallskip

From now on,  by a left inverse we will always mean a positive 
and normalized left inverse.\smallskip

An object $u\in\rZ^1(\sR_{K(M)})$  is said to be \textit{simple} whenever 
\begin{equation}
\label{Dc:1}
\eps(u,u) = \chi(u)\cdot 1_{u\otimes u}\ , \mbox{ where }\chi(u)\in\{1,-1\}\ .
\end{equation} 
If $u$ is simple, it turns out that
$y^u_x(o):\cR^\minperp(x)\to\cR^\minperp(x)$ 
is an \emph{automorphism} for any $o\in\Si_0(K(x^\minperp))$ (cfr. 
\cite[Proposition 4.12, Theorem 4.22]{Ruz3}).
If we denote the inverse of ${y}^u_x(o)$ by  ${y}^{u}_x(o)^{-1}$ it is easily seen that 
\begin{equation}
\label{Dc:1a}
 \phi_{z_1,z_2}(t)_o\doteq  y^{u}_x(o)^{-1}(t_o)\ , \qquad t\in(u\otimes z_1\ ,
 u\otimes z_2)\ ,  
\end{equation}
where $x$ is a point of $M$ causally disjoint from the closure of $o$, 
is a faithful left inverse of $u$. So any simple object
of $\rZ^1(\sR_{K(M)})$ has faithful left inverses. \smallskip

Denote by $\widetilde{\rZ}^1(\sR_{K(M)})$ the full
$\rC^*$--subcategory of $\rZ^1(\sR_{K(M)})$ whose objects have 
faithful left inverses. By applying   formulas
(A.1) (A.2) and (A.3) in \cite{Ruz1}, it easily follows that this category 
is closed under tensor product, direct sum, subobjects and
equivalence. Furthermore, this category is not trivial.
In fact, as observed above, any simple object of $\rZ^1(\sR_{K(M)})$
belongs to this category.\smallskip

Since the category $\widetilde\rZ^1(\sR_{K(M)})$ has left 
inverses and since it is closed under tensor product, direct sums 
and subobjects, it is possible to apply
the mathematical machinery of DHR analysis to define and 
classify the objects with finite statistics (see references 
quoted at the beginning of Section \ref{D}). An object $z$ of  
$\widetilde\rZ^1(\sR_{K(M)})$ has \emph{finite statistics} 
if it admits a standard left inverse $\phi$, that  is 
\[
\phi_{z,z}(\eps(z,z))^2 = c\cdot 1_z\ , \qquad c >0\ . 
\]
Let $\rZ^1(\sR_{K(M)})_{\mathrm{f}}$ be  the 
full $\rC^*$--subcategory of $\widetilde\rZ^1(\sR_{K(M)})$
whose objects with finite statistics. Then, 
$\rZ^1(\sR_{K(M)})_{\mathrm{f}}$ is closed under tensor product, 
direct sum and subobjects. Any object of this category 
is a finite direct sum of irreducible objects with finite
statistics.  Given  an irreducible 
object $z$ of $\rZ^1(\sR_{K(M)})_{\mathrm{f}}$ 
and a left inverse $\phi$, then 
\[
\phi_{z,z}(\eps(z,z)) = \la(z)\cdot 1_z\ , 
\]
where $\la(z)$ is an invariant of the 
equivalence class of $z$, called 
the \textit{statistics parameter}, and it is the product of two invariants: 
\[
\la(z) = \kappa(z)\cdot d(z)^{-1} \ \mbox{ where } \ \kappa(z)\in\{1,-1\}\ , \ \ 
  d(z)\in\mathbb{N}\ .
\] 
The possible statistics of $z$ are classified by the 
\textit{statistical phase} $\kappa(z)$ distinguishing para-Bose $(1)$ 
and para-Fermi $(-1)$ statistics and by the \textit{statistical dimension}
$d(z)$ giving the order of the para-statistics. Ordinary 
Bose and Fermi statistics correspond to $d(z)=1$.

In a symmetric tensor $\rC^*$-category 
an object $z$ has \emph{conjugates} if there exists 
an object $\overline{z}$ and a pair of arrows 
$r\in (\io,\overline{z}\otimes z)$ and 
$\overline{r}\in (\io,z\otimes \overline{z})$  satisfying the 
\emph{conjugate} equations
\begin{equation}
\label{Dc:3}
\overline{r}^*\otimes 1_z \cdot   1_z\otimes r = 1_z\ , \ \ \ \ 
r^*\otimes 1_{\overline{z}} \cdot    1_{\overline{z}}
\otimes \overline{r} = 1_{\overline{z}}\ .
\end{equation}
It is a well known fact that if an object has conjugates, then it has 
a faithful left inverse and finite statistics.  
So any object of $z$ having conjugates belongs to 
$\rZ^1(\sR_{K(M)})_{\mathrm{f}}$. We now show that any object of this category 
has conjugates. To this end it is enough to prove that simple objects 
have conjugates. So, consider a simple object $u$. Define 
\begin{equation}
\label{Dc:4}
\con{u}(b)  \doteq   y^{u}_x(\partial_0b)^{-1}(u(b)^*)\ , \qquad  b\in\Si_1(K(M))\ ,
\end{equation}
for some $x$ with $x\perp cl(|b|)$. Again, one first observe that 
the definition is independent of the choice of the point 
$x$. 
Finally,  one checks that 
$(u\otimes \overline{u})(b)=  \mathbbm{1}$ and that 
$(\overline{u}\otimes u)(b)   =  \mathbbm{1}$.
So, if we take  $r=\overline{r}=\mathbbm{1}$, then 
$r$ and $\overline{r}$ satisfy  the conjugate equations for 
$u$ and $\overline{u}$, cfr.\cite{Ruz3}. Then, 
the  above observation leads to the following conclusion.
\begin{theorem}
\label{Dc:6}
Any object of $\rZ^1(\sR_{K(M)})_{\mathrm{f}}$  has conjugates.
\end{theorem}
%


%
\section{The topological content}
\label{Y}

The twofold information contained in 1--cocycles, 
the charge and the topological content, can be splitted. 
We shall see that any 1--cocycle $z$ can be written as a suitable
composition (the join) of two 1--cocycles: the charge component
$\dec{z}$,  a topologically trivial 1--cocycle having the same charge quantum
numbers as $z$; the topological component $\chi_z$, a 1--cocycle 
that carries  the same representation 
of the fundamental group of $M$ as $z$ but it 
does not take values in the observable net.  This decomposition 
holds for any 1--cocycle of $\rZ^1(\sR_{K(M)})$. When we specialize 
to the finite statistics case, we shall find a relation between 
the statistics and the topological content of 1--cocycles.
This relation shall lead  us to discover a new invariant: the
topological dimension.\smallskip  

In order to decompose 1--cocycles into 
charge and topological component, we introduce the notion 
of path-frame which assigns to any 0--simplex a path-coordinate 
with respect to a fixed 0--simplex, the pole.
To be precise 
we fix a 0--simplex $o$, \emph{the pole}. For any 0--simplex $a$, 
we pick a path $p_{(a,o)}$ from $o$ to $a$ such that 
$p_{(o,o)}$ is homotopic to the trivial loop over $o$, i.e., 
the degenerate 0--simplex $\si_0o$. We
call the collection $P_o\doteq\{p_{(a,o)}\,|\,a\in\Si_0(K(M))\}$ 
a \emph{path-frame} with pole $o$.   
The \emph{translation} of a path-frame $P_o$ is the path-frame 
$P_{o}*o_1$ whose elements, denoted by $p_{(a,o_1)}$, are of the form 
$p_{(a,o)}*\overline{p_{(o_1,o)}}$. Note that
the translation  $P_{o}*o_1*o$ can be identified with 
$P_o$ since they have homotopic elements.


\subsection{Splitting} 
\label{Ya}

We now  show the splitting of a 1--cocycle 
into  charge and topological component.\bigskip

Fix a path-frame $P_o$ with pole $o$.
Given a 1--cocycle $z$  of $\rZ^1(\sR_{K(M)})$ define 
\begin{equation}
\label{Ya:1}
\dec{z}(b) \doteq z(p_{(\partial_0b,o)}*\overline{p_{(\partial_1b,o)}})\ , 
                      \qquad b\in\Si_1(K(M))\ .
\end{equation}
%
We call $\dec{z}$ the \emph{charge component} of $z$.\smallskip

It is very  easy to see that $\dec{z}$ is a topologically trivial 
1--cocycle, i.e., $\dec{z}\in\rZ^1_t(\sR_{K(M)})$. In fact, 
it follows straightforwardly from the definition that 
$\dec{z}$ is a 1--coboundary of $\rZ^1(K(M),\mathfrak{B}(\cH_0))$.
Moreover, given a 1--simplex $b$  for any 0--simplex $a$ with 
$|b|\perp a$ by (\ref{Ca:5}) we have 
$\dec{z}(b)\in \cR(a)'$. Thus $\dec{z}(b)\in\cR(|b|)$ by Haag duality.

We now  show that definition (\ref{Ya:1}) is independent, 
up to equivalence, of  the choice of the path-frame and of the choice
of the pole. Given another path-frame $Q_o$ define 
$s_a\doteq z(q_{(a,o)}*\overline{p_{(a,o)}})$ for any 0--simplex $a$. 
Then  
\[
s_{\partial_0b}\, z(p_{(\partial_0b,o)}*\overline{p_{(\partial_1b,o)}}) = 
z(q_{(\partial_0b,o)}*\overline{p_{(\partial_1b,o)}}) = 
z(q_{(\partial_0b,o)}*\overline{q_{(\partial_1b,o)}})\, s_{\partial_1b}\ ,
\]
for any 1--simplex $b$. Furthermore, since 
$q_{(a,o)}*\overline{p_{(a,o)}}$ is a loop over $a$, 
then $s_a\in\cR(a)$ because of the localization of the 
fundamental group. Thus, different choices of path-frames with the
same pole lead to
equivalent charge components of $z$. Now,  consider  another pole
$o_1$ and  the translation 
$P_{o}*o_1$ of $P_o$. Since 1--cocycles are homotopic 
invariant we have 
\begin{equation}
\label{Ya:1a}
z(p_{(\partial_0b,o)}*\overline{p_{(\partial_1b,o)}}) 
= z(p_{(\partial_0b,o_1)}*\overline{p_{(\partial_1b,o_1)}})\ , 
\end{equation}
because the paths $p_{(\partial_0b,o)}*\overline{p_{(\partial_1b,o)}}$
and  $p_{(\partial_0b,o)}*\overline{p_{(o,o_1)}}* p_{(o,o_1)}*
\overline{p_{(\partial_1b,o)}}$
$=p_{(\partial_0b,o_1)}*\overline{p_{(\partial_1b,o_1)}}$ 
are homotopic. 
This completes the proof of our claim.\smallskip 

We now go deep inside the relation between $z$ and its charge 
component. We fix a path-frame $P_o$. The first important 
observation  is that the morphisms of stalks associated with 
$z$ and $\dec{z}$ are equal. According to (\ref{Da:4}) and (\ref{Da:2}), 
it is enough to see that 
given a path $q$ and a 0--simplex $a$, with $|q|\perp a$,
and $A\in\cR(a)$, then 
\[
z(q)\, A \, z(q)^*  = z(
p_{(\partial_0q,o)}*\overline{p_{(\partial_1q,o)}}) \, A 
z(p_{(\partial_0q,o)}*\overline{p_{(\partial_1q,o)}})^* = 
\dec{z}(q)\, A\,  \dec{z}(q)^*\ ,
\]
where we have applied (\ref{Ca:6}) since the paths 
$q$ and $p_{(\partial_0q,o)}*\overline{p_{(\partial_1q,o)}}$ satisfy
the hypotheses of that relation. 
\begin{lemma}
\label{Ya:2}
The mapping $\cP_{\mathrm{c}}:\rZ^1(\sR_{K(M)})\to \rZ^1_t(\sR_{K(M)})$, which 
sends an object $z$ to its charged component ${\dec{z}}$, with respect
to a fixed path-frame $P_o$, and acts 
as the identity on  arrows $t\to t$, 
defines a faithful and  symmetric, covariant $^*$-functor. 
\end{lemma}
\begin{proof}
It is easily seen that $\cP_{\mathrm{c}}$ a faithful and covariant $^*$--functor. 
We only observe that if $t\in(z,z_1)$, then 
\[
t_{\partial_0b}\,\dec{z}(b) = 
t_{\partial_0b}\,z(p_{(\partial_0b,o)}*\overline{p_{(\partial_1b,o)}}) = 
z_1(p_{(\partial_0b,o)}*\overline{p_{(\partial_1b,o)}})\, t_{\partial_1b}= 
\dec{z_1}(b)\, t_{\partial_1b}\ ,
\]
for any 1--simplex $b$. We now prove that $\cP_{\mathrm{c}}$ preserves 
the tensor product.  Given a 1--simplex $b$,  pick $x\in M$ such that 
$|b|\in K(x^\minperp)$. Moreover,  pick pole $o_1$ 
in $K(x^\minperp)$. Given 
$z,z_1$, and using (\ref{Ya:1a}) we have 
\begin{align*}
\cP_{\mathrm{c}}(z\otimes z_1)(b) & =  \dec{z\otimes z_1}(b) \\
    & = z\otimes z_1\, (p_{(\partial_0b,o)}*\overline{p_{(\partial_1b,o)}}) \\
    & = z\otimes z_1\,(p_{(\partial_0b,o_1)}*\overline{p_{(\partial_1b,o_1)}}) \ ,
\end{align*}
where $p_{(a,o_1)}$ is the path associated with the translation 
$P_o*o_1$.  
Since $|b|,o_1$ are in  $K(x^\minperp)$, there are  paths 
$q_0$ and $q_2$ in $K(x^\minperp)$ which are homotopic respectively 
to $p_{(\partial_0b,o_1)}$ and $p_{(\partial_1b,o_1)}$ 
(Corollary \ref{Z:6}). The
previous equation and the homotopic invariance of 1--cocycles 
lead to $\cP_{\mathrm{c}}(z\otimes z_1)(b) = z\otimes z_1(q_0*q_1)$. 
Finally, by definition of the tensor product, by (\ref{Db:2a}) and by 
applying again homotopic invariance 
of 1--cocycles we have 
\begin{align*}
\cP_{\mathrm{c}}(z\otimes z_1)(b) & =  z\otimes z_1(q_0*q_1) = 
z(q_0*q_1)\times_x z(q_0*q_1) \\
& = z(p_{(\partial_0b,o_1)}*\overline{p_{(\partial_1b,o_1)}})
\times_x z(p_{(\partial_0b,o_1)}*\overline{p_{(\partial_1b,o_1)}}) \\
&      = \dec{z}(b)\times_x \dec{z_1}(b) \\
    & = \dec{z}\otimes \dec{z_1}\, (b)\ , 
\end{align*}
for any 1--simplex $b$. Note that we have 
used the fact that $z$ and $\dec{z}$ define
the same morphisms of stalks, as observed just before this lemma. 
Finally we prove that 
$\eps(z,z) = \eps(\dec{z},\dec{z})$. To this end, using the same 
notation in (\ref{Db:6}), recall that 
$\eps(z,z)_a$ does not depend on the choice of the paths  
$p$ and $q$ in (\ref{Db:6}). Now  the paths 
$p_{(\partial_0p,o)}*\overline{p_{(a,o)}}$ and 
$p_{(\partial_0q,o)}*\overline{p_{(a,o)}}$ have the same endpoints 
as $p$ and $q$ respectively. However we cannot 
replace in the definition of $\eps(z,z)_a$ the paths $p$ and $q$
by $p_{(\partial_0p,o)}*\overline{p_{(a,o)}}$  and 
$p_{(\partial_0q,o)}*\overline{p_{(a,o)}}$ respectively, since the
latter do not belong to  $K(x^\minperp)$ in
general. However, we note that by Corollary \ref{Z:6} (see the observation 
below the corollary) there are two paths $p_1$ and $q_1$ 
lying in $K(x^\minperp)$ which are homotopic to 
$p_{(\partial_0p,o)}*\overline{p_{(a,o)}}$  and 
$p_{(\partial_0q,o)}*\overline{p_{(a,o)}}$ respectively. 
So we have 
\[
\eps(z,z)_a  =  z(q)^*  \times_x z(p)^* \  z(p)  \times_x  z(q) \\
             =  z(q_1)^*  \times_x z(p_1)^* \  z(p_1)  \times_x
            z(q_1) \ .
\]
Observing (\ref{Db:1}) and 
applying homotopic invariance of 1--cocycles 
\begin{align*}
z(p_1) \times_x
            z(q_1) & = z(p_1)\,y^z_x(\partial_0p_1)(z(q_1)) \\
                   & =
                  z(p_{(\partial_0p,o)}*\overline{p_{(a,o)}})\,
         y^z_x(\partial_0p_1)(z(p_{(\partial_0q,o)}*\overline{p_{(a,o)}}))\\
          &    =     \dec{z}(p)\, y^z_x(\partial_0p)(\dec{z}(q))\\
          &    = \dec{z}(p)\, y^{\dec{z}}_x(\partial_0p)(\dec{z}(q))\\
          &    = \dec{z}(p)\times_x \dec{z}(q)\ ,
\end{align*}
where we have used the identities 
$z(p_{(\partial_0p,o)}*\overline{p_{(a,o)}})=\dec{z}(p)$ 
$z(p_{(\partial_0q,o)}*\overline{p_{(a,o)}})=\dec{z}(q)$, which derive 
from (\ref{Ya:1}),  and that $z$ and $\dec{z}$ define the same 
endomorphisms of stalks.
\end{proof}
\begin{remark}
\label{Ya:2a}
Two observations on $\cP_{\mathrm{c}}$ are in order. 
First, it easily follows from the  definition that 
$\cP_{\mathrm{c}}$ is a projection,
i.e. $\cP_{\mathrm{c}}\cP_{\mathrm{c}}=\cP_{\mathrm{c}}$. Secondly,
the functor $\cP_{\mathrm{c}}$  is not full in general. In fact,  
assume that $z$ carries a non-trivial representation of the fundamental 
group. Take $\ell$ a loop over $o$, define 
$t_a\doteq z(p_{(a,o)}*\ell*\overline{p_{(a,o)}})$ for any  
$a\in\Si_0(K(M))$. By the localization of the fundamental group 
$t_a\in\cR(a)$ for any 0--simplex. Moreover,
$t_{\partial_0b}\, \dec{z}(b)  =$ 
$z(p_{(\partial_0b,o)}*\ell*\overline{p_{(\partial_0b,o)}})\,
z(p_{(\partial_0b,o)}*\overline{p_{(\partial_1b,o)}})$ 
$ =
z(p_{(\partial_0b,o)})\, z(\ell) z(\overline{p_{(\partial_1b,o)}})$ 
$= \dec{z}(b)\, t_{\partial_1b}$. 
So, $t\in (\dec{z},\dec{z})$, while $t\not\in(z,z)$ in general. 
\end{remark}

We now introduce a second 1--cocycle encoding the topological 
content of $z$. Fix a path-frame $P_o$. 
Define
\begin{equation}
\label{Ya:3}
\chi_z(b) \doteq z(\overline{p_{(\partial_0b,\,o)}}*b*p_{(\partial_1b,\,o)})\ ,
                      \qquad 
                      b\in\Si_1(K(M))\ .
\end{equation}
We call $\chi_z$ the \emph{topological component} of $z$.\smallskip

The topological component $\chi_z$ is a 1--cocycle of $K(M)$ taking values in 
$\cR(o)$. In fact, by  the localization of the fundamental 
group,  $\chi_z$ takes values in $\cR(o)$.
Moreover
\begin{align*}
\chi_z(\partial_0c)\, \chi_z(\partial_2c) & = 
 z(\overline{p_{(\partial_{00}c,o)}}*\partial_0c*p_{(\partial_{10}c,o)}) \,
    z(\overline{p_{(\partial_{02}c,o)}}*\partial_2c*p_{(\partial_{12}c,o)}) \\
& =  z(\overline{p_{(\partial_{01}c,o)}})\, z(\partial_0c)\, 
    z(p_{(\partial_{02}c,o)}) \,
    z(\overline{p_{(\partial_{02}c,o)}}) z(\partial_2c) 
    z(p_{(\partial_{11}c,o)})\\
& =  z(\overline{p_{(\partial_{01}c,o)}})\, z(\partial_1c) 
     z(p_{(\partial_{11}c,o)})\\
& = \chi_z(\partial_1c)\ ,
\end{align*}
for any  2--simplex $c$. Hence $\chi_z$ is a 1--cocycle of 
the category $\rZ^1(K(M),\cR(o))$.
We now observe that
$z$ and $\chi_z$ contains the same topological information, namely 
\begin{equation}
\label{Ya:4}
 \chi_z(\ell) = z(\ell)\ , \qquad \ell\in\mathrm{Loops}_{K(M)}(o)
\end{equation}
In words  $z$ and $\chi_z$ define the same representation of the
first homotopy group. In fact assume that $\ell$ is of the form
$\ell=b_n*\cdots *b_1$. Then 
\begin{align*}
  \chi_z(\ell) & = z(\overline{p_{(\partial_0b_n,o)}}*
                   b_n*p_{(\partial_1b_n,o)})\, 
                   z(\overline{p_{(\partial_0b_{n-1},o)}}*
                   b_n*p_{(\partial_1b_{n-1},o)})\, 
                \cdots \\ 
               & \phantom{=} \ \ \ \cdots \qquad 
               z(\overline{p_{(\partial_0 b_2,o)}}*b_2*
                 p_{(\partial_1 b_2,o)})\, 
               z(\overline{p_{(\partial_0b_1,o)}}*b_1*p_{(\partial_1b_1,o)})\\
  & = z(\overline{p_{(\partial_0b_n,o)}})\, z(b_n) \,z(b_{n-1}) 
      z(p_{(\partial_1b_{n-1},o)})\, 
               \cdots \,\\
          &  \phantom{=} \ \ \ \cdots \qquad   z(\overline{p_{(\partial_0b_2,o)}})\, z(b_2)
               z(b_1) \, z(p_{(\partial_1b_1,o)})\\
 & = z(\overline{\si_0o})\, z(\ell)\, z(\si_0o)\\
  & = z(\ell)\ .
\end{align*}
Note that  equation (\ref{Ya:4}) says
that the representation of the fundamental group 
carried by the topological component of a 1--cocycle 
depends neither on the choice of the path-frame nor on the choice of the pole.
\begin{remark}
\label{Ya:4a}
We point out the geometrical meaning 
of  the topological component of a 1--cocycle $z$ of $\rZ^1(\sR_{(K(M)})$.  
We recall that  1--cocycles of a poset taking values in a group  
are flat connections of the poset 
(see \cite{RR07,RRV}). So the topological component 
$\chi_z$ is a \emph{holonomy} 
of the flat connection $z$. We also note that 
the definition of $\chi_z$ is the same as the definition of the 
reduced connection  in the Ambrose-Singer theorem for posets 
\cite[Theorem 4.28]{RR07}. 
\end{remark}
%
Further informations about the relation between  $z$ and its topological 
and charge components will be obtained by means of the following 
embedding theorem.
\begin{theorem}
\label{Ya:5}
Given $z\in\rZ^1(\sR_{K(M)})$, fix a path-frame $P_o$ and, 
for any $X\in\cR^z(M,o)$,  define 
\begin{equation}
\label{Ya:6}
 \varrho_a(X)\doteq z(p_{(a,o)})\, X\,z(p_{(a,o)})^*\ , \qquad a\in\Si_0(K)\ .
\end{equation}
Denote the family of mappings $X\to \varrho_a(X)$, $a\in\Si_0(K(M))$,
by $\varrho$. Then 
\begin{itemize}
\item[(i)] $\varrho:\cR^z(M,o)\to (\dec{z},\dec{z})$ 
          is a injective $^*$--morphism;
\item[(ii)] $\varrho:\cZ(\cR^z(M,o))\to 
          (z,z)$,
\end{itemize} 
where $\cZ(\cR^z(M,o))\doteq\cR^z(M,o)\cap \cR^z(M,o)'$, the centre 
of $\cR^z(M,o)$.   
\end{theorem}
\begin{proof}
Let us start by observing that $\varrho_a(X)\in\cR(a)$.
To this end, let us consider a loop $\ell$ over $o$. Observe that 
$\varrho_a(z(\ell))= z(p_{(a,o)}*\ell*\overline{p_{(a,o)}})$, and 
that $p_{(a,o)}*\ell*\overline{p_{(a,o)}}$ is a loop over $a$. 
Then $\varrho_a(z(\ell))$ belongs to the von Neumann algebra 
$\cR^z(M,a)$. Since the unitaries $z(\ell)$ generate $\cR^z(M,o)$ 
and since $\varrho_a$ is, clearly,  normal we have 
$\varrho_a(X)\in \cR^z(M,a)$. By Theorem \ref{Ca:4} we have 
$\varrho_a(X)\in \cR(a)$. \smallskip

$(i)$  Given a 1--simplex $b$ we have  
\begin{align*}
\varrho_{\partial_0b}(X)\, \dec{z}(b) & = 
 z(p_{(\partial_0b,o)})\, X \,z(p_{(\partial_0b,o)})^*\, 
   z(p_{(\partial_0b,o)})\, z(p_{(\partial_1b,o)})^* \\
 & = z(p_{(\partial_0b,o)})\, X\, z(p_{(\partial_1b,o)})^* \\
 & = \dec{z}(b) \, \varrho_{\partial_0b}(X)\ .
\end{align*}
Thus $\varrho_{\partial_0b}(X)\in (\dec{z},\dec{z})$.\smallskip

$(ii)$
Assume that $X$ belongs to the centre of $\cR^z(M,o)$. Given a
1--simplex $b$ we have 
\begin{align*}
\varrho_{\partial_0b}(X)\, z(b) & = 
z(p_{(\partial_0b,o)})\, X \, 
                     z(p_{(\partial_0b,o)})^*\, z(b) \\
& =  z(p_{(\partial_0b,o)})\, X \, 
                     z(\overline{p_{(\partial_0b,o)}}*b*p_{(\partial_1b,o)})\,
                     z(p_{(\partial_1b,o)})^* \\
& =  z(p_{(\partial_0b,o)})\,
                     z(\overline{p_{(\partial_0b,o)}}*b*p_{(\partial_1b,o)})\,
                     X \,  z(p_{(\partial_1b,o)})^* \\
& =  z(b)\, z(p_{(\partial_1b,o)})\,X \,                    
                     z(p_{(\partial_1b,o)})^* \\
& = z(b)\,\varrho_{\partial_0b}(X)\ ,
\end{align*}
because $\overline{p_{(\partial_0b,o)}}*b*p_{(\partial_1b,o)}$ 
is a loop over  $o$.
\end{proof}
%

The first application of Theorem \ref{Ya:5} derives from the following 
observation. Given a 1--cocycle $z$, define  
$(z,z)_a\doteq \{t_a \, |\, t\in (z,z)\}$ as the component $a$ 
of the set of intertwiners of the 1--cocycle. It is easily seen 
that any component of the algebra of intertwiners 
is a von Neumann algebra and that, since 
the poset is pathwise connected, this algebra is 
isomorphic to the full algebra of intertwiners.  Now, 
since, by definition, $p_{(o,o)}$ is homotopic to $\si_0o$ we have 
$\varrho(X)_o= X$ for any $X\in\cR^z(M,o)$. Then by Theorem \ref{Ya:5}
we have 
\begin{eqnarray}
 \cR^z(M,o)      & \subseteq  & (\dec{z},\dec{z})_o\ , \label{Ya:7} \\
 \cZ(\cR^z(M,o)) & \subseteq & (z,z)_o\ . \label{Ya:8}
\end{eqnarray}
The next result is a direct consequence of (\ref{Ya:8}).
\begin{corollary}
\label{Ya:9}
Let $z$ be a 1--cocycle of $\rZ^1(\sR_{K(M)})$. 
If $z$ is irreducible, then 
the representation of $\pi_1(M,o)$ associated with $z$ 
             is a factor representation, i.e., 
             $\cZ(\cR^z(M,o))= \mathbb{C}\mathbbm{1}$.
\end{corollary}
%
%
\subsection{Joining}
\label{Yb}
We learned that  a 1--cocycle $z$ splits in a pair $(\chi_z,\dec{z})$ of 
1--cocycles:  
$\dec{z}$ is topologically trivial but contains the charge structure 
of $z$ (this will be clearly shown in Subsection \ref{Yc});
$\chi_z$ encodes  the topological content of the $z$. 
We now show that two such 1--cocycles can be  joined 
together to form a 1--cocycle of $\rZ^1(\sR_{K(M)})$.\smallskip

Consider a topologically trivial 
1--cocycle $z\in\rZ^1_t(\sR_{K(M)})$, and 
a 1--cocycle $\varphi$ of the category $\rZ^1\big(K(M),\cR(o)\big)$.
We say that $\varphi$ and $z$ are \emph{joinable} whenever 
$\varphi(b)\in (z,z)_o$ for any 1--simplex $b$, 
and define the \emph{join} of $\varphi$ and  $z$, with respect to a
path-frame $P_o$, as 
\begin{equation}
\label{Yb:1}
 (\varphi\Join z)(b) \doteq  z(b)\, z(p_{(\partial_1b,o)})\,\varphi(b)\, 
                             z(p_{(\partial_1b,o)})^*, 
\ \qquad b\in\Si_1(K(M))\ . 
\end{equation}
Our aim is to show that the join $\varphi\Join z$ is 1--cocycle 
of $\rZ^1(\sR_{K(M)})$, whose topological and charge component
are equivalent to $\varphi$ and $z$, respectively, 
and that any 1--cocycle of $\rZ^1(\sR_{K(M)})$  arises as the join 
of its topological component and its charge component (note  
that, because of (\ref{Ya:7}), these are joinable).\smallskip 
  
We start by showing a property of the join. 

%
%
\begin{lemma}
\label{Yb:2a}
Fix a path-frame $P_o$. Let $z\in\rZ^1_t(\sR_{K(M)})$ and 
$\varphi\in \rZ^1\big(K(M),\cR(o)\big)$ be joinable. Then 
\[
(\varphi\Join z)(q) =  z(p_{(\partial_0q,o)})\,\varphi(q)\, 
                             z(p_{(\partial_1q,o)})^*\ ,
\]
for any path $q$.
\end{lemma}
\begin{proof}
We start by observing 
that a 1--cocycle $z$ is topologically trivial if, and only if,  
it is path-independent, that is, 
$z(p) = z(\tilde p)$ for any pair $p$,  $\tilde p$ of paths 
with the same endpoint. With this in mind, 
note that the above formula holds for paths which are formed by a single 
1--simplex. By induction, assume that 
the formula holds for paths formed by $n$ 1--simplices. Let 
$q$ be a path formed by $n$ 1--simplices and consider the path 
$q_1=q*b$.  
Then 
\begin{align*}
(\varphi  \Join z) (q_1) & =  
(\varphi\Join z)(q)\, (\varphi\Join z)(b) \\
& =  z(q)\, z(p_{(\partial_1q,o)})\,\varphi(q)\, 
              z(p_{(\partial_1q,o)})^*\, z(b)\, z(p_{(\partial_1b,o)})\,
\varphi(b)\, z(p_{(\partial_1b,o)})^*\\
& =  z(q)\, z(p_{(\partial_1q,o)})\,\varphi(q)\, 
              z(\overline{p_{(\partial_1q,o)}}*b*p_{(\partial_1b,o)})\,
\varphi(b)\, z(p_{(\partial_1b,o)})^*\\
& =  z(q)\, z(p_{(\partial_1q,o)})\,\varphi(q)\,
\varphi(b)\, z(p_{(\partial_1b,o)})^*\\
& =  z(p_{(\partial_0q,o)})\,\varphi(q*b)\, 
   z(p_{(\partial_1b,o)})^*\\
& =  z(p_{(\partial_0q_1,o)})\,\varphi(q_1)\, 
   z(p_{(\partial_1q_1,o)})^*\ ,
\end{align*}
where we have used topological triviality of $z$ ($z$ gives   
the same value on paths having  the same endpoints).
\end{proof}
Note that as a consequence of the above lemma,
the join does not depend on the choice of the path-frame. 
This follows directly from  
the  formula in the statement of Lemma \ref{Yb:2a} and from 
the topological triviality of the 1--cocycle $z$.\smallskip 

We now are in a position to prove the main result of this section.
\begin{theorem}
\label{Yb:3}
Fix a path-frame $P_o$. Let $z\in \rZ^1_t(\sR_{K(M)})$ and 
$\varphi\in\rZ^1\big(K(M),\cR(o)\big)$ be joinable. 
Then the following assertions hold.
\begin{itemize}
\item[(i)] The join $\varphi\Join z$ is a 1--cocycle of
$\rZ^1(\sR_{K(M)})$ whose topological and charge components are 
equivalent to $\varphi$ and $z$, respectively.
\item[(ii)] Any 1--cocycle $z$ of $\rZ^1(\sR_{K(M)})$ is 
the join $\chi_z\Join \dec{z}$, with respect to $P_o$, of its 
topological component $\chi_z$ 
with its charged component $\dec{z}$.
\end{itemize}
\end{theorem}
\begin{proof}
$(i)$ Clearly by Lemma \ref{Yb:2a} the join satisfies the 1--cocycle 
identity. 
Moreover $\varphi\Join z$ is localized. 
In fact take a 1--simplex $b$ which lies 
in $K(x^\minperp)$ and let $a$ be a 0--simplex in 
$K(x^\minperp)$ such that $a\perp |b|$. Then 
\begin{align*}
(\varphi\Join z)(b)\, A & = 
 z(b)\, z(p_{(\partial_{1}b,o)})\, \varphi(b)\, 
   z(p_{(\partial_{1}b,o)})^* \, A\\ 
& = 
 z(b)\, z(p_{(\partial_{1}b,o)})\, \varphi(b)\, 
   y^z_x(o)(A)\,  z(p_{(\partial_{1}b,o)})^*\\
& =  z(b)\, z(p_{(\partial_{1}b,o)})\, 
     y^z_x(o)(A)\,  \varphi(b)\, z(p_{(\partial_{1}b,o)})^*\\ 
& =  z(b)\, 
     y^z_x(\partial_1b)(A)\,  z(p_{(\partial_{1}b,o)})\, \varphi(b)\, z(p_{(\partial_{1}b,o)})^*\\ 
& =  z(b)\, 
     y^z_x(\partial_1b)(A)\,  z(p_{(\partial_{1}b,o)})\, \varphi(b)\, z(p_{(\partial_{1}b,o)})^*\\ 
& = y^z_x(\partial_0b)(A)\,   z(b)\, z(p_{(\partial_{1}b,o)})\, \varphi(b)\, z(p_{(\partial_{1}b,o)})^*\\ 
& = A\,   (\varphi\Join z)(b)\ , 
\end{align*}
where we have used the properties of localized transportable 
endomorphisms of stalks and the fact that $\chi\in (z,z)_o$. 
By Haag duality $(\varphi\Join z)(b)\in \cR(|b|)$; hence 
$\varphi\Join z\in\rZ^1(\sR_{K(M)})$. We now prove that the 
topological component $\chi_{\varphi\Join z}$ is equivalent to 
$\varphi$. First of all, 
given a loop $\ell =b_n*\cdots*b_1$ over $o$, observe that 
\begin{align*}
\chi_{\varphi\Join z}(\ell) & =
 (\varphi\Join z)(\ell) \\
& = z(b_n)\, z(p_{(\partial_{1}b_n,o)})\, \varphi(b_n)\, 
   z(\overline{p_{(\partial_{1}b_n,o)}}*b_{n-1}*
     p_{(\partial_{1}b_{n-1},o)})\,\varphi(b_{n-1})\cdots   \\ 
 & \phantom{=}\cdots \varphi(b_2)\,z(\overline{p_{(\partial_{1}b_2,o)}}*b_1*p_{(\partial_{1}b_1,o)})\, \varphi(b_1)\, 
   z(\overline{p_{(\partial_{1}b_1,o)}})\\ 
& = z(b_n*p_{(\partial_{1}b_n,o)})\, 
     \varphi(b_n)\, \varphi(b_{n-1})\,\cdots \,\varphi(b_{2})\,\varphi(b_{1})
   z(p_{(\partial_{1}b_1,o)})\\
& = z(p_{(o,o)})\, 
     \varphi(\ell) 
   z(p_{(o,o)})\\
& =      \varphi(\ell)\ ,  
\end{align*}
because $z(\overline{p_{(\partial_{1}b_{k+1},o)}}*b_{k}*
     p_{(\partial_{1}b_{k},o)})=\mathbbm{1}$, with $k=1,\ldots, n-1$,  
since $z$ is topologically trivial. So define 
$s_a\doteq  \varphi(p_{(a,o)})$ for any 0--simplex $a$. Then 
     $s_a\in\cR(a)$ and by the above identity we have 
\begin{align*}
 s_{\partial_0b}\, \chi_{\varphi\Join z}(b) & =
 \varphi(p_{(\partial_0b,o)})\, 
 (\varphi\Join z)(\overline{p_{(\partial_0b,o)}}*b*p_{(\partial_1b,o)})\,\\
& =  \varphi(p_{(\partial_0b,o)})\, 
 \varphi(\overline{p_{(\partial_0b,o)}}*b*p_{(\partial_1b,o)})\\ 
& =  \varphi(b)\, \varphi(p_{(\partial_1b,o)})\\
& =  \varphi(b)\, s_{\partial_1b} \ ,
\end{align*}
for any 1--simplex  $b$; thus $\chi_{\varphi\Join z}$ is equivalent 
to $\varphi$ in $\rZ^1(K(M),\cR(o))$.
We prove that 
$\dec{\varphi\Join z}$ is equivalent to $z$ in 
$\rZ^1_t(\sR_{K(M)})$. Define 
$t_a\doteq z(p_{(a,o)})\, \varphi(p_{(a,o)})^*\, z(p_{(a,o)})^*$,   
for any 0--simplex  $a$.
We first observe that $t_a\in\cR(a)$. In fact, since the topological 
component of the join is $\varphi$ we have that 
$\cR^{\varphi\Join z}(o) = \cR^{\varphi}(o)$. Now, by Lemma \ref{Ya:2a} 
and using topological triviality of $z$ we have 
\begin{align*}
\varrho^{\varphi\Join z}_a(\varphi(p_{(a,o)})^*) &  
= (\varphi\Join z)(p_{(a,o)})\, \varphi(p_{(a,o)})^* \, 
(\varphi\Join z)(\overline{p_{(a,o)}})\\
& = z(p_{(a,o)})   \varphi(p_{(a,o)}) \, z(p_{(o,o)})^*  
 \varphi(p_{(a,o)})^* \,  z(p_{(o,o)}) \,  
 \varphi(p_{(a,o)})^*\, z(p_{(a,o)})^*\\ 
& = z(p_{(a,o)})   \,\varphi(p_{(a,o)})^*\, z(p_{(a,o)})^*\\
& = t_a\ ,
\end{align*}
for any 0--simplex $a$, where $\varrho^{\varphi\Join z}$  is the embedding
(\ref{Ya:6}) associated with the join. The preceding observation 
and  Theorem \ref{Ya:5} imply that $t_a\in\cR(a)$. 
According to (\ref{Ya:1}) and by Lemma \ref{Yb:2a} we have 
\begin{align*}
t_{\partial_0b}\, \dec{\varphi\Join z}(b) & = 
t_{\partial_0b}\, (\varphi\Join 
z) (p_{(\partial_0b,o)}*\overline{p_{(\partial_1b,o)}}) \\
& = t_{\partial_0b}\,z(p_{(\partial_0b,o)})
  \varphi(p_{(\partial_0b,o)}*\overline{p_{(\partial_1b,o)}})\,  
 z(p_{(\partial_1b,o)})^* \\
& =  z(p_{(\partial_0b,o)})\, \varphi(p_{(\partial_0b,o)})^*\,
       \varphi(p_{(\partial_0b,o)}*\overline{p_{(\partial_1b,o)}})\,  
 z(p_{(\partial_1b,o)})^* \\
& =  z(p_{(\partial_0b,o)})\, 
       \varphi(p_{(\partial_1b,o)})^*\, 
 z(p_{(\partial_1b,o)})^* \\
& =  z(p_{(\partial_0b,o)})\, z(p_{(\partial_1b,o)})^*\, 
     z(p_{(\partial_1b,o)})\,   \varphi(p_{(\partial_1b,o)})^*\, 
 z(p_{(\partial_1b,o)})^*\\
& =  z(b) \, t_{\partial_1b}\ ,
\end{align*}
for any 1--simplex $b$, and this proves the equivalence.\\
\indent $(ii)$ As already observed if  $z\in\rZ^1(\sR_{K(M)})$, then 
$\chi_z$ and $\dec{z}$ are joinable.
Then 
\begin{align*}
(\chi_z\Join \dec{z})(b) &  
       = \dec{z}(b)\, \dec{z}(p_{(\partial_1b,o)})\, \chi_z(b)\,  
       \dec{z}(\overline{p_{(\partial_1b,o)}}) \\
& =  z(p_{(\partial_0b,o)}*\overline{p_{(\partial_1b,o)}})\,  
     z(p_{(\partial_1b,o)})\, 
     z(\overline{p_{(\partial_0b,o)}}* b* p_{(\partial_1b,o)})\, 
      z(p_{(\partial_1b,o)})^*\\
& =  z(b)\ ,
\end{align*}
for any 1--simplex $b$, and this completes the proof.
\end{proof}
As an easy consequence of this theorem, we have the following 
\begin{corollary}
\label{Yb:4}
Assume that the fundamental group of $M$ is Abelian. 
Fix a path-frame $P_o$. Then for any 
irreducible 1--cocycle $z$ we have that $\chi_z\in\rZ^1(K(M),\mathbb{C})$
and  
\[
z(b)= \chi_z(b)\, \dec{z}(b)\ , \qquad b\in\Si_1(K(M))\ . 
\]
\end{corollary}
\begin{proof}
Since $\pi_1(M,o)$ is Abelian the algebra
$\cR^z(M,o)$ is Abelian. Hence  $\cR^z(M,o)=\mathbb{C}\mathbbm{1}$
because of Corollary \ref{Ya:9}$(ii)$. The proof now follows from 
Theorem \ref{Yb:3}.
\end{proof}

\subsection{The topological dimension}
\label{Yc}
We now focus on irreducible 1--cocycles having finite statistics. 
We shall see that the charge component completely 
encodes the  charge structure of 1--cocycles and find  
a relation between the  statistical dimension 
and the topological  content  of 1--cocycles: 
the representation of the fundamental group 
carried by the 1--cocycle is, up to infinite multiplicity, 
irreducible and finite dimensional. The dimension of this
representation is bounded from above by the statistical dimension, 
and is a new invariant of sectors: the topological dimension.

\begin{proposition}
\label{Yc:1}
Let $z$ be an irreducible object of $\rZ^1(\sR_{K(M)})_{\mathrm{f}}$
whose statistical parameter $\la(z)$ is equal to $\kappa(z)\, d(z)^{-1}$. 
Fix a path-frame $P_o$. Then the following assertions hold.
\begin{itemize}
\item[(i)] If $r$ and $\overline{r}$ solve the conjugate equations 
            for $z$ and $\overline{z}$, then the same arrows 
            solve the conjugate equations 
            for $\dec{z}$ and $\dec{\overline{z}}$.
\item[(ii)] $\dec{z}$ is an object with finite statistics 
           which is, in general,  a finite direct sum
           $\dec{z}=z_1\oplus\cdots \oplus z_m$, with $m\leq d(z)$,  
           of irreducible objects of $\rZ^1_t(\sR_{K(M)})_{\mathrm{f}}$ 
           having the same statistical phase as $z$ and whose  statistical 
           dimension $d(z_i)$  satisfies  $d(z)=d(z_1)+\cdots +d(z_m)$. 
\end{itemize}
\end{proposition}
\begin{proof}
$(i)$ The mapping 
$\cP_{\mathrm{c}}:\rZ^1(\sR_{K(M)})\to
\rZ^1_t(\sR_{K(M)})$, sending an object $z$ into its charge
component $\dec{z}$, with respect to $P_o$, and acting as the identity 
on arrows  is a faithful and symmetric tensor $^*$--functor
(Lemma \ref{Ya:2}). It follows straightforwardly from these properties
that if $r$ and $\overline{r}$ solve the conjugate equations 
            for $z$ and $\overline{z}$, then 
$\cP_{\mathrm{c}}(r)=r$ and
$\cP_{\mathrm{c}}(\overline{r})=\overline{r}$ solve the conjugate 
equations for $\dec{z}$ and $\dec{\overline{z}}$. This, in particular, 
implies that  $\dec{z}$ has finite statistics. \\
\indent $(ii)$ 
Given $z',z''\in \rZ^1_t(\sR_{K(M)})_\mathrm{f}$, 
define 
\begin{equation}
\label{Yc:2}
 \psi_{z',z''}(X)\doteq 
\frac{1}{d(z)}  \, \cP_{\mathrm{c}}(r)^*\otimes 1_{z''} 
             \cdot 1_{\cP_{\mathrm{c}}(\overline{z})}\otimes X\cdot 
                               \cP_{\mathrm{c}}(r)\otimes 1_{z'}\ ,
\end{equation}
with $X\in(\cP_{\mathrm{c}}(z)\otimes z', \cP_{\mathrm{c}}(z)\otimes
z'')$\footnote{Note that  
$\psi_{z',z''}(X)$ 
$= d(z)^{-1}  \, r^*\otimes 1_{z''} 
             \cdot 1_{\dec{\overline{z}}}\otimes X\cdot 
                               r\otimes 1_{z'}$ 
with $X\in(\dec{z}\otimes z', \dec{z}\otimes z'')$,
according to the definition of $\cP_{\mathrm{c}}$.}. By 
\cite[Prop.4.5]{LR},  the collection $\psi\doteq\{\psi_{z',z''} \, | \, 
z',z''\in \rZ^1_t(\sR_{K(M)})_\mathrm{f}\}$, defines 
a standard left inverse  of $\dec{z}$, within the category 
$\rZ^1_t(\sR_{K(M)})_\mathrm{f}$. Moreover $\dec{z}$ has the same 
statistical dimension as $z$. In addition, since $\cP_{\mathrm{c}}$ 
is symmetric we have 
\begin{align*}
\psi_{\dec{z},\dec{z}}(\eps(\dec{z},\dec{z})) & = 
 d(z)^{-1} \,  \cP_{\mathrm{c}}(r)^* \otimes 1_{\dec{z}} 
               \cdot 1_{\cP_{\mathrm{c}}(\overline{z})}\otimes 
                \eps(\dec{z},\dec{z})\cdot 
                \cP_{\mathrm{c}}(r)\otimes 1_{\dec{z}}\\
&= d(z)^{-1} \,  \cP_{\mathrm{c}}(r^* \otimes 1_{z}) 
               \cdot \cP_{\mathrm{c}}(1_{\overline{z}}\otimes 
                \eps(z,z)\cdot 
                \cP_{\mathrm{c}}(r\otimes 1_z)\\
&   = d(z)^{-1} \,  
 \cP_{\mathrm{c}}\big(r^* \otimes 1_z  \cdot 1_{\overline{z}}\otimes 
                     \eps(z,z)\cdot r^*\otimes 1_z \big)\\
&   =  \cP_{\mathrm{c}}(\la(z)\, 1_z)\\
&   = \la(z)\,  1_{\dec{z}}\ .
\end{align*} 
It is a well known fact, see for instance \cite{DHR}, that 
this relation implies that $\dec{z}$ is a finite direct sums 
of irreducible object of $\rZ^1_t(\sR_{K(M)})_{\mathrm{f}}$
which have the same 
statistical phase as $z$ and the sum of their statistical dimensions
is equal to $d(z)$.
\end{proof}

\begin{proposition}
\label{Yc:3}
The following assertions hold for any irreducible 1--cocycle $z$ 
with finite statistics. 
\begin{itemize}
\item[(i)] Let $r,\overline{r}$ be  arrows solving 
the conjugate equations for $z$ and $\overline{z}$. Then 
the functional 
\begin{equation}
\label{Yc:4}
\omega^z_o(X)\doteq \frac{1}{d(z)}\, r^*_o\,
y^{\overline{z}}_x(o)(X)\,r_o\ , \qquad X\in\cR^z(M,o)\ ,
\end{equation}
is a normal and faithful tracial state of $\cR^z(M,o)$.
\item[(ii)] The algebra $\cR^z(M,o)$ is a type $I_m$ factor 
with $m\leq d(z)$.
\end{itemize}
\end{proposition}
\begin{proof}
Fix a path-frame $P_o$ and consider the charge component 
$\dec{z}$, with respect to $P_o$,  of $z$.\\
\indent $(i)$ We have seen in the proof of Proposition \ref{Yc:2} that 
$\dec{z}$ has a standard left inverse $\psi$ in 
$\rZ^1_t(\sR_{K(M)})$, defined by  equation (\ref{Yc:3}). 
This implies that $\psi_{\io,\io}$ is a faithful tracial state 
of the algebra $(\dec{z},\dec{z})$. Observe in particular that 
\[
 \psi_{\io,\io}(t)_o  = \frac{1}{d(z)}  \, 
           (r^*\otimes 1_{\io})_o 
             \,  (1_{\dec{\overline{z}}}\otimes t)_o
             (r\otimes 1_{\io})_o =
     \frac{1}{d(z)}  \, 
            r^*_o\, y^{\overline{z}}_x(o)(t_o)\, 
             r_o =  \omega^z_o(t_o)\ .
\]
%
Hence $\omega^z_o$ is a faithful tracial state of 
$(\dec{z},\dec{z})_o$ because so is 
$\psi_{\io,\io}\restriction (\dec{z},\dec{z})_o$. Now the proof
follows by (\ref{Ya:7}) and by 
observing that the morphisms of stalks of 1--cocycles 
are locally normal. \\
\indent $(ii)$ Note that  $(\dec{z},\dec{z})_o$ 
is a finite dimensional algebra having at most $d(z)$ 
minimal mutually orthogonal projection. According 
to the definition of a finite type $I$ factor (see \cite{KR}),
the proof follows 
by Corollary \ref{Ya:9} and by  (\ref{Ya:7}).
\end{proof}
On the ground of this result we can introduce the 
following notion. 
\begin{definition}
\label{Yc:5}
Given a 1--cocycle $z$ with finite 
statistics, the \textbf{topological dimension} $\tau(z)$ of $z$ is 
the dimension of the factor $\cR^z(M,o)$. 
\end{definition}
Let us see  which are the main properties 
of this new notion and its meaning. \emph{First}, the topological dimension 
is a quantum number of superselection sectors, i.e., 
it is an invariant of the
equivalence class of a 1--cocycle, since $\cR^z(M,o)$ is spatially 
equivalent to $\cR^{z_1}(M,o)$ whenever  
$z$ is equivalent to $z_1$. \emph{Secondly}, 
the topological dimension is bounded 
from above by the statistical dimension ($(ii)$ of Proposition \ref{Yc:3}). 
\emph{Thirdly},
the topological dimension and the tracial state defined 
in (\ref{Yc:4}) characterize the topological content of a 1--cocycle.
To explain this point we need a preliminary result which 
is an easy  consequence of Proposition \ref{Yc:3}.
\begin{corollary}
\label{Yc:6}
Let $z$ be an irreducible object with finite statistics.
Then:
\begin{itemize} 
\item[(i)] $z$, as a representation of $\pi_1(M,o)$,  is equivalent to 
a representation of the form $\mathbbm{1}_{\cH_0}\otimes\si_z$ where 
$\si_z$ is a $\tau(z)$-dimensional irreducible representation 
of $\pi_1(M,o)$;
\item[(ii)] the normalized character $\mathrm{c}_{\si_z}$ of $\si_z$
     satisfies the equation 
\[
 \mathrm{c}_{\si_z}([\ell])= \omega^z_o(z(\ell)) \ , \qquad \ell\in 
\mathrm{Loops}_{K(M)}(o) \ ,   
\]
where $\omega^z_o$ is the functional (\ref{Yc:4})
\end{itemize}
\end{corollary}
\begin{proof}
$(i)$  Since  $\cR^z(M,o)$ is a type $\rI_{\tau(z)}$ factor, there is 
a unitary operator $U:\Hil_0\to \Hil_0\otimes \rC^{\tau(z)}$ 
such that 
$\cR^z(M,o)\cong \mathbbm{1}_{\cH_0}\otimes \mathbb{M}_{\tau(z)}$. 
Then the representation $\si_z$ of $\pi_1(M,o)$ defined by 
\[
\mathbbm{1}_{\cH_0}\otimes \si_z([\ell])\doteq  
 U\,z(\ell) \,U^*   \qquad \ell\in \mathrm{Loops}_{K(M)}(o)\ ,
\]
is  irreducible,  because 
$\mathbbm{1}_{\cH_o}\otimes (\si_z)'' =$
$(\mathbbm{1}_{\cH_o}\otimes \si_z)'' =$  
$U\,\cR^z(M,o) \,U^* = 
 \mathbbm{1}\otimes M_{\tau(z)}$. \\
\indent $(ii)$ follows from $(i)$ and from uniqueness of the trace for
the algebra $\mathbb{M}_{\tau(z)}$. 
\end{proof}
%
%
Now, recall that  the   category 
of finite dimensional representations of a topological group 
is equivalent to the category of finite dimensional representations  
of its Bohr-compactification \cite[Prop.16.1.3]{Dix}. Accordingly,
finite dimensional representations of a topological 
group are classified by their characters, and this shows our claim. 
\emph{Fourthly}, the topological dimension is stable under conjugation, 
this is the content of the next result. 
\begin{lemma}
\label{Yc:7}
Let $z$ be an irreducible object with finite  statistics. Then 
$z$ and the conjugate $\overline{z}$ have the same topological dimension.
\end{lemma}
\begin{proof}
Since $z$ is irreducible, $\overline z$ is irreducible 
and $d(z)=d(\overline{z})$. Let $\si_z$ and $\si_{\overline z}$ 
be  the $\tau(z)$- and  
$\tau(\overline z)$-dimensional representations of $\pi_1(M,o)$ 
associated, respectively,  
with $z$ and $\overline z$ by Corollary \ref{Yc:6}, and let  
$\mathrm{c}_{\si_z}$ and  $\mathrm{c}_{\si_{\overline z}}$ be the corresponding 
normalized characters. Given a loop $\ell$ over $o$, 
using equation (\ref{Db:8}) we have 
\begin{align*}
d(z) \ \omega^z_o(z(\ell)) & = r^*_o\ 
 \overline{y}(o)(z(\ell))\ r_o 
 =   r^*_o\  \overline{y}(o)(z(\ell))\
\overline{z}(\ell)\  \overline{z}(\ell)^*\  r_o \\
& =    r^*_o\ (\overline{z}\times z)(\ell)\
          \overline{z}(\ell)^*\ r_o 
 =    r^*_o\
          \overline{z}(\ell)^*\ r_o \\
& =    r^*_o\  \eps(z,\overline{z})_o\
                            y(o)(\overline{z}(\ell)^*)\
                            \eps(\overline{z},z)_o\  r_o 
 =    \overline{r}^*_o\
                            y(o)(\overline{z}(\ell)^*)\ \overline{r}_o 
\\ 
& =  d(\overline{z})\ \omega^{\overline{z}}_o(\overline{z}(\ell)^*)\ ,
\end{align*}
%
%
where we have used the relation that $\overline{r}=
\kappa(z) \cdot \eps(\overline{z},z)\cdot r$ (see \cite{LR}).
Hence 
\begin{equation}
\label{Yc:8} 
\omega^{\overline{z}}_o(\overline{z}(\ell)) =  
\omega^{z}_o(z(\ell))^*\ ,\qquad \ell\in\mathrm{Loops}_{K(M)}(o) \ .
\end{equation}
This relation and  $(ii)$ of Corollary \ref{Yc:6} imply  that 
$\mathrm{c}_{\si_{\overline{z}}}$  is equal to the adjoint  of  
$\mathrm{c}_{\si_z}$. Hence 
$\si_{\overline{z}}$ is equivalent to the conjugated representation 
of $\si_z$; the latter is irreducible and has dimension $\tau(z)$.
\end{proof}

\begin{remark}
\label{Yc:8a}
We do not define the topological dimension for reducible 
objects. This can be understood by the following observation.
If  $z$ is an irreducible 
1--cocycle with finite statistics, then 
the representation of the fundamental group associated 
with $z\oplus z$ and $z$ are equivalent. In fact by Corollary
\ref{Yc:6} we have 
$(\mathbbm{1}_{\cH_0}\otimes \si_{z}) \oplus 
(\mathbbm{1}_{\cH_0}\otimes \si_{z}) \cong \mathbbm{1}_{\cH_0}\otimes
\si_z$, 
because of the infinite multiplicity. Thus, 
it is not possible to extend the topological dimension 
to reducible objects by additivity. 
\end{remark}
Finally, we show the structure of those 1--cocycles 
whose topological dimension equals the statistical dimension.  
\begin{lemma}
\label{Yc:9}
Let $z$ be an irreducible object with finite statistics. 
Assume that $\tau(z)=d(z)$. 
\begin{itemize}
\item[(i)] Then $\dec{z} =u^{\oplus{\tau(z)}}$ the $\tau(z)$--fold   
direct sum of  a simple object $u$ of $\rZ^1_t(\sR(K(M))$. 
\item[(ii)] If  $d(z)=1$ then $\tau(z)=1$, so 
$\dec{z}$ is a simple object, while $\chi_z$ takes values 
in $\mathbb{C}$; hence
\[
 z(b) = \chi_z(b)\,\dec{z}(b)\ , \qquad b\in\Si_1(K(M))\ .
\]
\end{itemize}
\end{lemma}
\begin{proof}
$(i)$ If $\tau(z)=d(z)$, then  $(\dec{z},\dec{z})$
has $d(z)$  mutually orthogonal projections.
Since $\dec{z}$ has statistical dimension equal to 
$d(z)$ and since the statistical dimension is additive,
$\dec{z}$ is a finite direct sum of $d(z)$ simple subobjects
$u_i$. Moreover, all the subobjects of $\dec{z}$ are equivalent. 
Note that $\cR^z(M,o)$ is type $I_{\tau(z)}$ factor. So 
as a linear vector space it has dimension $\tau(z)^2$. 
Observe that if there were a subobject $u_1$, to say, 
which is not equivalent to the other subobjects of $\dec{z}$,
then the algebra $(\dec{z},\dec{z})_o$ would have, as linear vector space,
dimension less than $(\tau(z)-1)^2+ 1$. This leads to a contradiction 
because $\cR^z(M,o)\subseteq (\dec{z},\dec{z})_o$.
$(ii)$ follows from $(i)$.
\end{proof}

\section{Existence and physical interpretation}
\label{F}

The first aim of  this section is to show that for any irreducible 
finite dimensional representation of the fundamental group 
of the spacetime $M$, there is an irreducible 
object with finite statistics carrying, 
up to infinite multiplicity, this representation. This says 
that the category $\rZ^1(\sR_{K(M)})_\mathrm{f}$ 
describes the Bohr-compactification of the fundamental group 
of the spacetime. The second aim is to show how the topology 
of spacetime affects the 
charges associated with $\rZ^1(\sR_{K(M)})_\mathrm{f}$ 
and to point out the analogy with the 
Ehrenberg-Siday-Aharonov-Bohm effect. \bigskip

Recall that any representation of the fundamental group of the
spacetime $M$ defines, up to equivalence, a unique representation 
of the fundamental group of $K(M)$.  Thus,  let $\si$ be an irreducible 
$n$--dimensional representation of $\pi_1(K(M),o)$. 
Now, pick a  
a simple object $u$, possibly $u=\io$, of $\rZ^1_t(\sR_{K(M)})$ 
with say  Bose-statistics. Define 
$z\doteq u^{\oplus n}$ and note that 
the algebra $(z,z)_o$ is spatially equivalent to
$\mathbbm{1}_{\cH_0}\otimes \mathbb{M}_n$. Let $U:\cH_0\to \cH_0\otimes \mathbb{C}^n$ 
be the unitary operator yielding this equivalence. Define 
\[
\widetilde\si([\ell])\doteq U^*\, \mathbbm{1}_{\cH_0}\otimes \si([\ell])\, U\ , \qquad 
[\ell]\in \pi_1(K(M),o)\ .
\]
Observe that $\widetilde\si$ is a factor representation taking values
in $(z,z)_o$.  Following \cite{Ruz3}, by this representation 
we can define a 1--cocycle $\varphi_\si$ of $K(M)$
taking values in $(z,z)_o$. The representation 
of $\pi_1(K(M),o)$ associated with $\varphi_\si$ is equal to $\widetilde\si$.
Now, define  
\[
z_\si\doteq \varphi_\si\Join z\ .
\]
By Theorem \ref{Yb:3} $z_\si$ is a 1--cocycle of $\rZ^1(\sR_{K(M)})$
such that the representation of the fundamental group associated 
with $z_\si$ is $\widetilde\sigma$ 
(see within the proof the cited theorem).
Hence, this fact and Theorem \ref{Ca:8} prove the existence of 
topologically non-trivial net representations for 
the observable net $\sA_{K(M)}$.\smallskip

\indent We now show that $z_\si$ is an irreducible object 
with Bose-statistics and 
statistical dimension equal to $n$. We need a preliminary observation. 
Since  $\chi_{z_\si}$ is equivalent to $\varphi_\si$ and 
$\dec{z_\si}\cong z$ (Theorem \ref{Yb:3}),  without loss of
generality, from now on 
we assume $\chi_{z_\si}=\varphi_\si$ and $\dec{z_\si}= z$.
We now prove that $z_\si$ is irreducible. 
Assume that there is a non-zero projection 
$t\in (z_\si,z_\si)$ such that $t\ne \mathbbm{1}$. Clearly 
$t\in(z,z)$ and $t_o$ commutes with 
$\varphi_\si$. Moreover, $U\, t_o \, U^*$ is a non-zero 
projection of $1_{\cH_0}\otimes \mathbb{M}_n$ which is 
different from the identity of this algebra and commutes with $\si$.
This leads to a contradiction because $\si$ is irreducible.
Hence $z_\si$ is irreducible. \\ 
\indent Concerning the statistics,  one can deduce 
by the particular form of $z$, that there 
is an isometry $t\in (v,z^{\otimes n})$ 
such that $t\cdot t^*=\mathrm{a}^n_{z}$,
where $\mathrm{a}^n_{z}$ is the  
totally antisymmetric projection of 
$(z^{\otimes n},z^{\otimes n})$ 
defined by the representation $\eps^n_{z}$ of the permutation group 
$\mathbb{P}(n)$ associated with $z$, and  
$v$ is  a Bosonic simple   subobject $z^{\otimes n}$ (see
\cite{DHR}). Now, we observe that as a consequence 
of Lemma \ref{Ya:2} the representation $\eps^n_z$ 
equals the representation $\eps^n_{z_\si}$ for any $n$ 
(note that 
the functor $\cP_\mathrm{c}$ in Lemma \ref{Ya:2} 
acts as the identity on arrows). 
This,  in particular,  implies 
that  $\mathrm{a}^n_{z_\si} =\mathrm{a}^n_{z}$ 
and $\eps({z}_\si^{\otimes n}, {z}_\si^{\otimes n})= 
      \eps(z^{\otimes n}, z^{\otimes n})$. \\
\indent On these premises, 
using the defining properties of a permutation symmetry,
since $v$ is a Bosonic simple object,  we
have  $\mathbbm{1} = \eps(v,v) = t^*\otimes t^*\cdot 
\eps(z^{\otimes n}, z^{\otimes n})\cdot  t\otimes t$, 
which is equivalent to saying that 
$\mathrm{a}^n_{z}\otimes \mathrm{a}^n_{z} 
   =  \mathrm{a}^n_{z}\otimes \mathrm{a}^n_{z}\cdot  
\eps(z^{\otimes n}, z^{\otimes n})$.  Hence 
$\mathrm{a}^n_{z_\si}\otimes \mathrm{a}^n_{z_\si} 
=  \mathrm{a}^n_{z_\si}\otimes \mathrm{a}^n_{z_\si}\cdot  
\eps(z_\si^{\otimes n}, z_\si^{\otimes n})$. 
Accordingly, the subobject  $w$  of $z_\si^{\otimes n}$  associated with  
the projection $\mathrm{a}^n_{z_\si}$ is a simple object. 
This is enough 
to prove that $z_\si$ has finite statistics \cite{Ruz1}. By
Proposition \ref{Yc:1}
$z_\si$ is Bosonic and has statistical dimension 
equal to $n$. \\
\indent This leads to the following existence theorem.
\begin{theorem}
\label{F:1}
Let $\si$ be an irreducible and  $n$-dimensional representation
of the fundamental group of the spacetime $M$. Then,  there  
exists a 1--cocycle $z$ of $\rZ^1(\sR_{K(M)})$ 
having the following properties
\begin{itemize}
\item[(i)] $z$ is an irreducible object with finite statistics whose 
    statistical and topological dimension are equal to $n$;
\item[(ii)] the representation of the fundamenal group of the spacetime $M$
     associated with $z$ is equivalent to 
     $\mathbbm{1}_{\cH_0}\otimes \si$.
\end{itemize}
\end{theorem}
We draw on some consequences of this result. 
First, we point out that 
Theorem \ref{F:1} implies the existence of topologically 
non-trivial superselection sectors even when the only DHR-sector of
the net  $\sR_{K(M)}$ is the vacuum, i.e.,  $\rZ^1_t(\sR_{K(M)})$ 
is formed by finite direct sums of the trivial 1--cocycle $\io$. 
Secondly,  since any finite dimensional representation 
of $\pi_1(M,o)$ appears, up to infinite multiplicity, 
as the representation of $\pi_1(M,o)$  associated with 
a 1--cocycle with finite statistics,  
what we have shown is that the topological content of the category 
$\rZ^1(\sR_{K(M)})_{\mathrm{f}}$ is the Bohr-compactification 
of the fundamental group of $M$.\bigskip

We now turn to explain how the topology of the spacetime $M$ affects,
if not trivial, the charges  $\rZ^1(\sR_{K(M)})_{\mathrm{f}}$. 
Consider an irreducible  1--cocycle $z$  with finite statistics. 
Fix a point $x$ of the spacetime, 
and consider the family $y^z_x(o)$, with $o\in K(x^\minperp)$, 
of localized transportable endomorphisms of the stalk
$\cR(x^\minperp)$ associated with $z$. According to the interpretation given 
in DHR analysis, $y^z_x(o)$ is a representation describing 
a charge within the region $o$; the 1--cocycle 
$z$ is the transporter of this charge (see Remark \ref{Da:6a}). Then, 
transport this charge 
from $o$ to another region $\tilde o$
along two different paths $p$ and $q$ such that 
the loop $p*q$ over $o$ is not homotopic to the trivial loop.
Then 
\[
 z(p*\overline q)\,y^z_x(o) =  z(p)\, y^z_x(\tilde o) z(\overline q)
 = y^z_x(o) z(p*\overline q) \ne y^z_x(o)\ .
\]
This means that, analogously to the  
Ehrenberg-Siday-Aharonov-Bohm effect \cite{ES,AB}
the final state differs from the initial one of the
unitary $z(p*\overline q)$. 
The analogy is actually tighter if one thinks 
of 1--cocycles  as flat connections of a principal bundle 
over the poset $K(M)$ \cite{RR07,RRV}). 
Then  the difference 
from the initial to the final state is the parallel transport 
of the flat connection $z$ along the loop $p*\overline{q}$.\smallskip


\section{Comments and outlook} 
The present paper is, in our opinion, 
an interesting contribution to the topic of 
the existence of quantum effects induced by the topology of spacetimes. 
We have shown the existence
of a new class of superselection sectors having 
well defined charge and
topological contents, in the case in which the spacetime is multiply 
connected. These sectors are sharply localized in the same sense 
of those discovered by Doplicher, Haag and Roberts \cite{DHR}, but 
they are  affected by  the spacetime topology in a way similar 
to the quantum geometric phases.  
In a certain sense, the results of the present paper bring some support  to the idea, suggested by Ashtekar and Sen 
\cite{AS}, that non-trivial topologies may induce the existence 
of a new kind of particles (see also \cite{Hol}). 
We think that the main and the new 
contribution given by the present paper to that idea 
resides in the fact that the exposed results are model-independent 
and based on a few, physically reasonable, assumptions. \smallskip

We have shown the charge and the topological content  
of sectors of $\rZ^1(\sR_{K(M)})$ in a fixed spacetime 
background $M$. As said at the beginning,  
it would be interesting to understand 
the locally covariant \cite{BFV} behavior of these sectors. 
This may clarify some issues in the analysis of 
the locally covariant structure of DHR-sectors 
\cite{BR}.\smallskip

We now point out a central question arising from our results.\smallskip

\noindent \emph{Is there an underlying gauge theory giving rise 
to the charged sectors of 
$\rZ^1(\sR_{K(M)})_{\mathrm{f}}$?} \smallskip

We are asking  whether it is possible either to provide models of gauge fields giving rise to 
sectors $\rZ^1(\sR_{K(M)})_{\mathrm{f}}$  and, in general, 
 whether one is able to reconstruct  the fields  and the gauge 
group underlying the charges $\rZ^1(\sR_{K(M)})_{\mathrm{f}}$,
as  Doplicher and Roberts have shown to happen for DHR-sectors and 
for BF-sectors (the Buchholz-Fredenhagen charges \cite{BF})  
in Minkowski spacetime \cite{DR90}.\\
\indent As far as models are concerned, a first positive, but very preliminary step has been provided in \cite{BFM}: there, it is proven that a  
a massive bosonic quantum field in a 2-dimensional spacetime (the Einstein cylinder) has a non-trivial topological cocycle that gives rise to non-trivial unitary representations of the fundamental group of the circle. Note that, on 2-dimensional Minkowski spacetime, the model does not have any DHR sector of the usual kind besides the vacuum \cite{Mug}. It is an intriguing question whether our selection criterion gives something really different from DHR, or new perspectives, in low dimensions. We hope to return elsewhere to this question too.
 
As far as the abstract construction is concerned, the question about Doplicher-Roberts reconstruction Theorem has, in our opinion, two different answers according to  whether 
the fundamental group of the spacetime is Abelian or not. 
In the Abelian case, we think 
that there are no important differences from the scenery suggested 
by the Doplicher-Roberts reconstruction: there should 
exist a field net acted upon globally by the gauge group 
(in this direction goes the example in \cite{BFM}). 
Conversely, in the non-Abelian case, we expect 
the scenery of the Doplicher-Roberts reconstruction to break down. 
It is reasonable to think that this happens  
because of the operation of join (\ref{Yb:1}) 
that couples the charge and the topological
component of sectors. When
the fundamental group is not Abelian the coupling 
between the topological and the charge component is, in general, 
\emph{local}: the function describing that coupling within Definition (\ref{Yb:1}) depends on 1--simplices.
Conversely, this coupling is \emph{global} in the Abelian case 
because this function reduces to the identity for any 1--simplex 
(Corollary \ref{Yb:4}).\smallskip

It is not clear what could be the mathematical 
structure of the field theory underlying the charges 
$\rZ^1(\sR_{K(M)})_{\mathrm{f}}$ in the non-Abelian case. 
One wonders whether it may resemble, at least in a certain sense, 
what people describes as topological field theories. 
Indeed, in that framework, theories of flat connections 
(for instance, Chern-Simons \cite{Freed}) are considered, as much as we do at the quantum level by considering 1-cocycles with values operators.\bigskip

\noindent{\textbf{Acknowledgement.} We gratefully acknowledge discussions 
with Klaus Fredenhagen and John E. Roberts. We wish to warmly thank  Miguel S\'anchez for sharing with us his insights in Lorentzian geometry.

\appendix 
\numberwithin{equation}{section}

\section{Proof of Theorem \ref{Ca:8}} 
\label{X}
We prove the equivalence between $\rS\rC(\sA_{K(M)})$ 
and $\rZ^1(\sR_{K(M)})$. \smallskip

Let $\{\pi,\psi\}$ and element of $\rS\rC(\sA_{K(M)})$. Given a
1--simplex $b$, take a simply connected subspacetime $N$ such that 
$cl(|b|)\subset N$, and  define 
\begin{equation}
\label{X:1}
  z^\pi(b)\doteq   W_a^{N\partial_0b} \, {W_a^{N\partial_1b}}^*, \qquad 
     cl(a)\subset N, \ a \perp |b| \ .   
\end{equation}
where $W^{N o}$ is the unitary satisfying the selection criterion (\ref{Bb:1}). 
First of all we prove this  definition    is well posed.
By property 2   of (\ref{Bb:1}), if $\tilde a\subseteq a$ then 
\[
   W_{\tilde a}^{N \partial_0b} \, {W_{\tilde a}^{N \partial_1b}}^* =
   W_{a}^{N\partial_0b} \, \psi_{a\tilde a } \,
                               \psi^*_{a\tilde a } 
{W_{a}^{N \partial_1b}}^* =
   W_{a}^{N  \partial_0b} \,
{W_{a}^{N\partial_1b}}^* \ .
\]
Since $cl(|b|)\subset N$ we have that $|b|$ is a diamond of $N$. So 
the causal complement of $|b|$, relatively to $N$, is pathwise connected.
This and  the above 
identity leads to the independence of the choice of $a$. 
Concerning the independence of the choice of $N$, consider 
a second simply connected subspacetime  $N'$ such that 
$cl(|b|)\subseteq N'\cap N$. By Lemma \ref{Z:4} there is 
a diamond $\cO$ and a diamond $a$ such that 
$\cO\subseteq  N'\cap N$, $cl(|b|),cl(a)\subset \cO$ and 
$|b|\perp a$. Note that $\cO$ is a simply connected subspacetime 
of $M$. Property 3 of (\ref{Bb:1}) implies that 
$W^{N\partial_ib}_a= W^{N'\partial_ib}_a= W^{\cO\partial_ib}_a$ for 
$i=0,1$. This and the independence of the choice of $a$, leads 
to the independence of the choice of $N$.
Hence $z^\pi$ is well defined. Now, the cocycle identity 
follows from the definition of $z^\pi$ and 
from the independence of the choice of $N$ and $a$. In fact given 
a 2--simplex $c$ take $N$  such that 
$cl(|c|)$. By Lemma \ref{Z:4} there is $a$ such that 
$cl(a)\subset N$ and $c\perp a$. Then 
\begin{align*}
 z^\pi(\partial_0c) z^\pi(\partial_2c) & =  
 W_a^{N\partial_{00}c} \, {W_a^{N\partial_{10}c}}^*\,
    W_a^{N \partial_{02}c} \, {W_a^{N\partial_{12}c}}^* \\
& =  W_a^{N \partial_{01}c} \, {W_a^{N\partial_{10}c}}^*\,
    W_a^{N \partial_{10}c} \, {W_a^{N\partial_{11}c}}^*\\
& =  W_a^{N\partial_{01}c} \,{W_a^{N\partial_{11}c}}^*\\
& = z^\pi(\partial_1c) \ .
\end{align*}
Concerning the locality condition, note that by 
outer regularity and Haag duality it is easily seen that 
$\cR(o) = \cap \{\cR(\tilde o)' \, | \, cl(o)\perp cl (\tilde o)\}$
(see for instance \cite{Ruz2}). So given a 1--simplex $b$ pick 
a 0--simplex $o$ such that $cl(|b|)\perp cl(o)$.
By the smoothability argument \cite{BS06} there is
a spacelike Cauchy surface $\cC$  
which contains the closure of the bases of the diamonds $|b|$ and
$o$. Moreover, since the closure of the bases  of $|b|$ and $o$
are disjoint there is a simply connected  open subset $G$ of $\cC$ which 
contains the closure of both the bases.
Then, the domain of dependence of $G$ 
is a simply connected subspacetime $N$ which contains $cl(|b|)$ and
$cl(o)$. Since (\ref{X:1})  is independent 
of the choice of $a$,  by property 1 of (\ref{Bb:1}) we have 
\[
 z^\pi(b)\,\io_o(A) = 
  W_o^{N\partial_0b} \, {W_o^{N\partial_1b}}^*\, \io_o(A)
 =   \io_o(A) \, W_o^{N\partial_0b} \, {W_o^{N\partial_1b}}^*
 = \io_o(A) \, z^\pi(b) \ ,
\]
for any $A\in\cA(o)$.
Since this holds for any $cl(o)\perp cl(|b|)$, the above 
observation implies that $z^\pi(b)\in \A(|b|)$.  This proves that 
$z^\pi\in\rZ^1(\sR_{K(M)})$.\smallskip

Consider now $\{\si,\phi\}\in\rS\rC(\sA_{K(M)}$ and an 
arrow $T\in(\{\pi,\psi\},\{\si,\phi\})$. For any 0--simplex $a$ 
take $o\perp a$ and $N$ such that $cl(a),cl(o)\subset N$. Define 
\begin{equation}
\label{X:2}
 t_{a} \doteq   V_o^{N\,a} \, T_o \, {W_o^{Na}}^* \ .
\end{equation}
where $W$ and $V$ are the unitary associated with $\{\pi,\psi\}$ and 
$\{\si,\phi\}$, respectively, by the selection criterion. 
As above $t$ is well defined and independent of the choices of
$o$ and $N$. Given a 1--simplex $b$, pick $N$ with
$cl(|b|)\subset N$ and let $o$ be such that  $|b|\perp o$ and 
$cl(o)\subset N$. Then 
\[
 t_{\partial_0b} z^\pi(b) 
=   V_o^{N\partial_0b} \, T_o \, {W_o^{N\partial_0b}}^*\,
    W_o^{N\partial_0b} \, {W_o^{N\partial_1b}}^* 
 = V_o^{N\partial_0b} \, T_o \, {W_o^{N\partial_1b}}^*
 = z^\si(b) \, t_{\partial_1b} \ .
\]
%
One can also easily see that $t_a\in\cR(a)$, hence 
$t\in(z^\pi,z^\si)$. \smallskip

We now construct the map from net cohomology to net representations 
satisfying the selection criterion. 
Given a 1--cocycle $z\in \rZ^1(\sR_{K(M)})$, define 
\begin{equation} 
\label{X:3}
\begin{array}{rcll}
\pi^z_a (A) &\doteq& z(q_a) \, \io_a(A)\,  z(q_a)^* \ , &  
a\in K(M)\ , 
A\in\cA(a) \ ,  \\
\psi^z_{a,\tilde a} & \doteq & z(a,\tilde a) \ , & \tilde a\subseteq a \ ,
\end{array}
\end{equation}
where $q_a$ is a path with $\partial_1q=a$ and $\partial_0q\perp a$, 
and $(a,\tilde a)$ denotes the  1--simplex 
such that $\partial_1(a,\tilde a)= \tilde a$, 
$\partial_0(a,\tilde a)= |(a,\tilde a)| =a$.
The independence of the chosen path $q_a$
is a consequence of (\ref{Ca:6}).  This also implies that 
$\pi^z_a  = z(a,\tilde a) \, \pi^z_{\tilde a } \, z(a,\tilde a)^*= 
         \psi^z_{a\tilde a} \, \pi^z_{\tilde a } \, {\psi^z_{a\tilde a}}^*$
for any $\tilde a \subseteq a$. Hence the pair  
$\{\pi^z,\psi^z\}$ is a net representation. Note that the 
1--cocycle $\zeta^{\pi^z}$, associated with this net 
representation by (\ref{A:6}),  is equivalent 
to $z$. In fact,  we have 
\[
\zeta^{\pi^z}(b)  = {\psi^z_{|b|\partial_0b}}^*\,\psi^z_{|b|\partial_1b}
                 =    z(|b|,\partial_0b)^*\, z(|b|,\partial_1b) 
                = z(b) \ . 
\]
for any 1--simplex $b$. We now prove that this net representation satisfies 
the selection criterion  (\ref{Bb:1}).
Given $ o\in K(M)$, let $N$ be simply connected subspacetime 
such that $cl(o)\subset N$.  Given 
$a\perp o$, with  $cl(a)\subset N$, define 
\begin{equation}
\label{X:4}
 W_a^{N o} \doteq z(p_{o,a}) \ ,
\end{equation}
where $p_{o,a}$ is the path from $a$ to $o$ whose support has 
closure contained in $N$. 
Clearly this definition does not depend on $p$ since $N$ is simply
connected. Furthermore, given a path $q_a$ as in definition 
(\ref{X:3}), by relation (\ref{Ca:5}) we have 
\begin{align*}
 W_a^{N o}\, \pi^z_a(A)   & = 
        z(p_{o,a})\, z(q_a) \, \io_a(A) \, z(q_a)^* \\
 &      =         z(p_{o,a}*q_a) \, \io_a(A) \, z(q_a)^* 
       =      \io_a(A)\,    z(p_{o,a}) \\
    & = \io_a(A)\,  W_a^{N o},
\end{align*}
for any $A\in\cA(a)$, 
because the endpoints of $p_{o,a}*q_a$ 
are causally disjoint from $a$. Hence $\{\pi^z,\psi^z\}$ is a sharp 
excitation of the reference representation.  Finally,  given 
$t\in(z,\hat z)$ define 
\begin{equation}
\label{X:5}
T_a\doteq t_a, \qquad a\in\Si_0(K(M)) \ .
\end{equation}
Given  $A\in\A(a)$ we have 
\begin{align*}
 T_a\, \pi^z_a(A)  & = t_a \, z(q_a) \, \io_a(A)  
                             \, z(q_a)^*
 = \hat{z}(q_a) \, t_{\partial_1q_a} \, \io_a(A)  \, 
   z(q)^* \\
& = \hat{z}(q_a) \, \io_a(A) \, t_{\partial_1q_a} \,  z(q_a)^* 
 = \hat{z}(q_a) \, \io_a(A) \, \hat{z}(q_a)^*  \, 
t_{a} \\
& = \pi^{\hat{z}}_a(A) \, T_a.
\end{align*}
Moreover, it is easily seen that 
$T_a\psi^z_{a,\tilde a}= \psi^{\hat{z}}_{a,\tilde a} T_{\tilde a}$, if
$\tilde a\leq a$. Thus, $T$ is an intertwiner from 
$\{\pi^z,\psi^z\}$ to $\{\pi^{\hat{z}},\psi^{\hat{z}}\}$.\smallskip  

Now, 
the functor $F:\rS\rC(\sA_{K(M)})\to \rZ^1(\sR_{K(M)})$ 
defined by means of  the equations 
(\ref{X:1}) and (\ref{X:2}): 
given  $\{\pi,\psi\},\{\si,\phi\}$ 
and $T\in \big(\{\pi,\psi\},\{\si,\phi\}\big)$ define 
\begin{equation}
\label{X:6}
\begin{array}{ll}
F(\{\pi,\psi\})(b) \doteq z^\pi(b)\ , & b\in\Si_1(K(M)) \ ,  \\[3pt]
F(T)_a \doteq   V_o^{Na} \, T_o \, {W_o^{Na}}^*, & a\in\Si_0(K(M)) \ , \\
\end{array}
\end{equation}
where $W$ and $V$ are the operators associated with 
$\{\pi,\psi\}$ and $\{\si,\phi\}$, respectively, by the selection criterion. 
The functor $G: \rZ^1(\sR_{K(M)})\to \rS\rC(\sA_{K(M)})$ is defined by  
equations (\ref{X:3}) and (\ref{X:4}): 
given $z,\hat z\in\rZ^1(\sR_{K(M)})$ and $t\in(z,\hat z)$,
define 
\begin{equation}
\label{X:7}
\begin{array}{l}
G(z) \doteq \{\pi^z,\psi^z\} \ , \\
G(t) \doteq   t \ .
\end{array}
\end{equation}
We first study the composition $FG$.
Consider a 1--simplex $b$. Pick  a $0$--simplex $a$ with 
$cl(a)\subset N$ and $a \perp |b|$.  By using 
(\ref{X:4}) and (\ref{X:1}) we have  
\[
  FG(z)(b) = 
  z(p_{\partial_0b,a})\,  z(p_{\partial_1b,a})^* = 
    z(p_{\partial_0b,a}*\overline{p_{\partial_1b,a}}) =   z(b) \ , 
\]
because $p_{\partial_0b,a}*\overline{p_{\partial_1b,a}}$ is a path
from $\partial_1b$ to $\partial_0b$ and its closure is in $N$.
Hence it is homotopic to $b$ because $N$ is simply 
connected. Moreover by (\ref{X:5}) (\ref{X:4}) (\ref{X:2}) we have
$FG(t)_a   
   =  \hat{z}(p_{a,o}) \, t_o  \, z( \tilde p_{a,o})^* 
     =  t_a$ for any $0$--simplex $a$,  
because  $p_{a,o}$ and $\tilde p_{a,o}$ are paths from $o$ to $a$. 
Hence $FG=\mathrm{id}_{\rZ^1(\sR_{K(M)})}$. We now show that 
there is a natural isomorphism 
from $GF$ and $\mathrm{id}_{\mathrm{SC}(\sA_{K(M)})}$. 
First of all, 
given $a$, take a simply connected region $N$ which contains 
$cl(a)$ and pick $o,\tilde a$ such that 
$o\perp a,\tilde a$ and $\tilde a \perp o$.
Define 
\begin{equation}
\label{X:8}
 S_a(\{\pi,\psi\})\doteq 
{W_a^{N\tilde a}}^*\, W_{o}^{N\tilde a}\, {W_{o}^{Na}}^*,
\end{equation}
where $W$ are the unitaries associated with $\{\pi,\psi\}$ by the
selection criterion.
By the definitions of $G$ and $F$ we have 
\[
 GF(\{\pi,\psi\})_a(A)  = 
F(\{\pi,\psi\})(q_a) \, \io_a(A)  \, F(\{\pi,\psi\})(q_a)^*
      = W_o^{N a} \, {W_o^{N\tilde a}}^*   \,\io_a(A)\, 
    W_o^{N\tilde a}\,{W_o^{N a}}^* \ , 
\]
where $\tilde a\perp a$. In fact, the path $q_a$   
is from $a$ to $a^{\minperp}$. Since, the definition of 
$F(\{\pi,\psi\})$ does depend 
neither on the choice of the path nor on the choice of the 0--simplex in 
$a^\minperp$, we have considered a  path which lies $N$ 
from $a$ to $\tilde a$. Finally since $N$ is simply connected 
$F(\{\pi,\psi\})(q_a)$ depends only on the endpoints of $q_a$. 
Thus  $F(\{\pi,\psi\})(q_a)=W_o^{N a} \, {W_o^{N\tilde a}}^*$. 
Using this and (\ref{X:8}) we have 
\begin{align*}
S_a(\{\pi,\psi\})\,  GF(\{\pi,\psi\})_a(A) & = 
   S_a(\{\pi,\psi\})\,  W_o^{N a} \, {W_o^{N\tilde a}}^*   \,\io_a(A)\, 
      W_o^{N\tilde a}\,{W_o^{N a}}^* \\
& = {W_a^{N \tilde a}}^*\, \io_a(A)\,  W_o^{N\tilde a}\,{W_o^{N a}}^* 
 = \pi_a(A)\, {W_a^{N \tilde a}}^*\, W_o^{N\tilde a}\,{W_o^{N a}}^* \\  
& = \pi_a(A) \, S_a(\{\pi,\psi\}) \ .
\end{align*}
Furthermore, given $a_1\subseteq a$ we have 
\begin{align*}
\psi_{aa_1}\, S_{a_1}(\{\pi,\psi\}) & = 
\psi_{aa_1}\, 
{W_{a_1}^{N\tilde a}}^*\, W_{o}^{N\tilde a}\, {W_{o}^{Na_1}}^* 
 = {W_{a}^{N\tilde a}}^*\, W_{o}^{N \tilde a}\, {W_{o}^{Na_1}}^*\\ 
& = {W_{a}^{N\tilde a}}^*\, W_{o}^{N\tilde a}\, {W_{o}^{Na}}^*\, 
W_{o}^{Na}\,{W_{o}^{Na_1}}^* \\
& = S_{a}(\pi) \, \psi^{z^\pi}_{aa_1} \ . 
\end{align*}
This proves that  $S(\{\pi,\psi\})$ is a unitary 
intertwiner from $GF(\{\pi,\psi\})$ to $\{\pi,\psi\}$.
Finally, given  $T\in(\{\si,\phi\},\{\pi,\psi\})$ then 
\begin{align*}
 S_a(\{\pi,\psi\})\,  GF(T)_a  & =  S_a(\{\pi,\psi\})\, F(T)_a \\
  & = {W_a^{N\tilde a}}^*\, W_{o}^{N \tilde a}\,
  {W_{o}^{N a}}^*\, W_o^{Na}\, T_o \, {V_o^{Na}}^*   
   = {W_a^{N\tilde a}}^*\, W_{o}^{N\tilde a}\, 
   T_o  \,{V_o^{N\tilde a}}^*\, V_o^{N\tilde a}\,
   {V_o^{Na}}^*  \\
  & = {W_a^{N\tilde a}}^*\, W_{a}^{N\tilde a}\, 
   T_a  \,{V_a^{N\tilde a}}^*\, V_o^{N\tilde a}\,
   {V_o^{Na}}^*  
  = T_a  \,{V_a^{N\tilde a}}^*\, V_o^{N\tilde a}\, 
   {V_o^{Na}}^* \\
 & = T_a  \,S_a(\{\si,\phi\}) \ ,
\end{align*}
where the property 
$W_{o}^{N\tilde a}\, 
   T_o  \,{V_o^{N\tilde a}}^* = W_{a}^{N\tilde a}\, 
   T_a  \,{V_a^{N\tilde a}}^*$ has been used; 
$W$ and $V$ denote the unitaries associated, respectively, with 
$\{\pi,\psi\}$ and $\{\si,\phi\}$ by the selection criterion.
This completes the proof of Theorem \ref{Ca:8}.

\section{Miscellanea on the causal structure}
\label{Z}
In this section we provide some results about the causal 
structure of  globally hyperbolic spacetimes. 
In what follows $M$ denotes a connected globally hyperbolic spacetime. 
References for this appendix are the same as those  quoted in 
Section \ref{Ba}. \smallskip

To begin with, we recall that if $A$ is a closed acrhonal set 
then the closure of $D^{\pm}(A)$ is the union of those points 
such that any inextendible past (forward) directed 
timelike curve starting from these points meets $A$.  
$D^{\pm}(A)$ denotes the future (past) domain of dependence 
of the set $A$ defined according to the convention  
of \cite{EH,One}.\smallskip

The first result is due to the courtesy of Miguel S\'anchez. We are indebted with him for the nice short proof.
\begin{lemma}[M. S\'anchez]
\label{Z:0}
Let $\cC$ be a spacelike Cauchy surface of $M$, and let $K$ be 
a nonempty closed subset of $\cC$. Then 
$D^{\pm}(K)$ is closed.
\end{lemma}
\begin{proof}
Let $q$ be a point in the boundary of  $D^+(K)$ not included in
$D^+(K)$. Then there is a past-directed inextendible causal curve through $q$ 
which does  not cross $K$. Nevertheless, 
it will cross $\cC$ at some point $p$. 
As $K$ is closed in $\cC$, a neighborhood $U$ of $p$ in $\cC$ does not 
intersect $K$. Easily, there are points in $U$ 
chronologically related to $q$.  
($p$ lies in
 $J^-(q)$, which is included in the adherence of $I^-(q)$). 
Let $p_1$ be one such point. 
The past directed  timelike curve from $q$ to $p_1$ cannot 
intersect $K$ (as cannot intersect $\cC$ again). 
This contradicts the fact that $cl(D^+(K))$ is
the set of points $x$ of $M$ such that every inextendible timelike curve
through $x$ crosses $K$.
\end{proof}
If $A\subseteq\cC$ let us denote by $int_\cC(A)$ the internal part of
$A$ in the relative topology of $\cC$. Note that as $\cC$ 
is a closed subset of $M$, the closure 
of $A$ in the relative topology of $\cC$ coincides 
with its  closure  in the topology of $M$.
\begin{lemma}
\label{Z:1}
Let $\cC$ be a spacelike Cauchy surface of $M$, and let $K$ be 
a nonempty closed subset of $\cC$. 
Assume that $K\subseteq cl(int_\cC(K))$. Then 
$cl(D^{\pm}(int_\cC(K)))=D^{\pm}(K)$.
 \end{lemma}
\begin{proof}
As far as the first inclusion is concerned, as  $D^+(int_\cC(K))\subset D^+(K)$, we have  
$cl(D^+(int_\cC(K)))$ $\subseteq D^+(K)$ by Lemma \ref{Z:0}.\\
\indent For the opposite inclusion, take $x\in D^+(K)$.  Note that 
if $x\in K$ then $x\in cl(D^+(int_\cC(K)))$. In fact as 
$K\subseteq cl(int_\cC(K))$ we have that $K\subset cl(D^+(int_\cC(K)))$. 
So assume that $x\in D^+(K)\setminus K$. 
Since, by Lemma \ref{Z:0}, $D^+(K)$  is closed, we have that  
$I^-(x)\cap \cC\subset K$.
However,  $I^-(x)$ is open in $M$, hence
$I^-(x)\cap \cC$ is open in the relative topology 
of $\cC$; thus  $I^-(x)\cap \cC\subseteq int_\cC(K)$.  
Take a sequence of points  $\{y_n\}$  converging to $x$ such that
$y_n\in I^-(x)\cap I^+(K)$ for any $n\in\mathbb{N}$.
Note that  $J^-(y_n)\subset I^-(x)$ because $y_n\in  I^-(x)$.
Hence $J^-(y_n)\cap \cC \subset I^-(x)\cap \cC \subseteq int_\cC(K)$. Thus
$y_n\in D^+(int_\cC(K))$ for any $n$. Since $y_n\to x$, then 
$x\in cl(D^+(int_\cC(K)))$. 
The same argument applies to the past development and the proof is over.
\end{proof}
\begin{corollary}
\label{Z:1a}
Let $\cC$ be a spacelike Cauchy surface of $M$, 
and let  $O$ be a nonempty open subset of $\cC$
in the relative topology of $\cC$. 
Assume that $int_\cC(cl(O))\subseteq O$. Then 
$cl(D^{\pm}(O))=D^{\pm}(cl(O))$.
\end{corollary}
\begin{proof}
Note that  the closed set $K\doteq cl(O)$ 
satisfies the hypothesis of Lemma \ref{Z:1}. 
In fact   $int_\cC(cl(O))$$= O$, because $O$ is open;
hence $cl(int_\cC(K))$$=cl(int_\cC(cl(O)))$$=cl(O)=K$, and the proof 
follows by Lemma \ref{Z:1}.
\end{proof}
This result applies to diamonds. Let  $o$ be a diamond 
whose base $G$ lies on a spacelike Cauchy surface $\cC$. 
$G$ is an open and relatively compact subset in the relative topology 
of  $\cC$. Moreover, $int_\cC(cl(G))=G$. Then as a consequence 
of the previous result we have 
\begin{equation}
\label{Z:2}
  cl(o) = cl(D(G)) = D(cl(G)) \ . 
\end{equation}
We now are in a position to prove the following property of diamonds.

\begin{lemma}
\label{Z:3}
Let $o$ be a diamond of $M$.  Then there is a sequence  $\{o_n, a_n\}$
of pairs of diamonds  based on the same Cauchy surface as $\cC$ and 
satisfying the following properties: 
\begin{itemize}
\item[(i)] $cl(o_{n+1}) \subset o_{n}$ for any
$n\in\mathbb{N}$ and $\cap_n o_{n}=cl(o)$;
\item[(ii)] $cl(a_{n})\subset o_{n}$ and 
           $cl(a_{n}) \perp cl(o_{n+1})$ for any $n\in\mathbb{N}$.
\end{itemize}
\end{lemma}
\begin{proof}
The proof is splitted in two parts. In the first step 
we construct, by induction, 
a sequence of pairs of ``formal'' diamonds 
satisfying all the properties in the statement. 
By a formal diamond we mean a  set satisfying all the properties 
to be a diamond  with the exception that it might be non relatively 
compact. In the second step we prove that from the above sequence
it is possible to extract a subsequence of diamonds. \smallskip

\emph{First step.}  According to the definition of diamonds there is 
a spacelike Cauchy surface $\cC$, 
a chart $(U,\phi)$ of $\cC$, and an open 
ball $B$ of $\mathbb{R}^3$ such that 
$cl(B)\subset \phi(U)$,  $o=D(G)$ where $G\doteq \phi^{-1}(B)$. 
Define $W_1\doteq U$. Since $cl(B)$ is a proper compact subset of 
the open $\phi(W_1)$ the distance $d$ of $B$ from 
$\mathbb{R}^3\setminus \phi(W_1)$ 
is strictly positive. Accordingly, denoting the radius of $B$ by  $r$,
define $B_1$ as the open ball having the same centre as $B$ 
and radius $r_1\doteq r+d/2$, and 
let $L_1$ be an open ball such that 
$cl(L_1) \cap cl(B) =\emptyset$ and $cl(L_1)\subset
B_1$. Observe that if we define 
$G_1\doteq \phi^{-1}(B_1)$ and $H_1\doteq \phi^{-1}(L_1)$, then 
by construction and by Lemma \ref{Z:1},  we have 
\[
cl(o),cl(D(H_1))\subset  D(G_1)\ , \ \ cl(o)\perp cl(D(H_1))\ , \ \ 
cl(G_1)\subset W_1 \ . \qquad 
 \qquad (*) 
\]
By induction,  for $n\geq 1$, assume that 
we have a 3-tuple  $(G_n,H_n, W_n)$ satisfying 
$(*)$. Define $W_{n+1}\doteq (W_n\setminus cl(H_n))\cap
G_{n}$. Applying the above construction with respect to 
$W_{n+1}$ we get a pair $G_{n+1}$ and $H_{n+1}$.  This procedure leads to a
sequence $\{G_n,H_n, W_{n}\}$ satisfying $(*)$ for any  $n$. Furthermore, 
by construction we have that $cl(H_{n})\perp cl(G_{n+1})$ and 
$cl(G_{n+1})\subset G_n$ for any $n$; moreover 
\[
\cap_n G_n = cl(G)\ . \qquad (**)
\]
Define $o_n\doteq D(G_n)$ and $a_n\doteq D(H_n)$. Hence we have 
$cl(o_{n+1})\subset o$; $cl(a_{n})\subset o_n$ and 
$cl(a_n)\perp cl(o_{n+1})$ for any $n$. We now show that 
$\cap_n o_n=cl(o)$.  Clearly $cl(o)\subseteq \cap_n o_n$. 
Let $x\in \cap_n o_n$. Since the sets $G_n$ belong 
to the same Cauchy surface $\cC$, there are two possibilities 
either $x\in D^{+}(G_n)$ for all $n$,  or    $x\in D^{-}(G_n)$ for all $n$.
Assume $x\in D^{+}(G_n)$ for all $n$. Then, 
$J^-(x)\cap \cC\subseteq G_n$ for all $n$; hence by $(**)$
$J^-(x)\cap \cC\subseteq cl(G)$, thus 
$x\in D^+(cl(G))=cl(D^+(G))\subset cl(o)$.   The same argument 
applies if $x\in D^{-}(G_n)$ for all $n$. This proves our claim.\smallskip

\emph{Second step.}  We now prove that 
the sequence $\{o_n,a_n\}$ is definitively formed by diamonds of $M$. 
To start with,  note that 
there is a smooth foliation $F:\Sigma\times \mathbb{R}\to M$  
of the spacetime, by spacelike Cauchy surfaces,   
such that $F(\Sigma,0)=\cC$ \cite{BS06}. To be precise, 
$F:\Sigma\times \mathbb{R}\to M$  is a diffeomorphism 
such that $F(\Sigma,t)$ is a spacelike Cauchy surface for any 
$t\in\mathbb{R}$ and the curve $\gamma_y(t)\doteq F(y,t)$ as 
$t$ varies in $\mathbb{R}$, for a fixed 
$y\in\Sigma$, is an inextendible forward directed timelike 
curve. The inverse function $F^{-1}$ is equal to 
$(h(x), \tau(x))$ where  
$h$ and $\tau$ are smooth functions from 
$M$ to $\Sigma$ and $\mathbb{R}$, respectively. On these grounds, 
since $cl(o)$ is compact there are $t_1,t_2\in\mathbb{R}$ such that 
$t_1 < t_2$ and  
\[
   t_1 < \{\tau(x)\in\mathbb{R}\,|\,  \ x\in cl(o)\}  < t_2\ ; 
\]
in other words  $cl(o)$ is a proper subset of 
$F(\Sigma,(t_1,t_2))$. Define 
Let $K^\Sigma \doteq \{ h(x)\in\Sigma \, |\, 
 x\in cl(G_1)\}$, where $G_1$ is the first set constructed above. 
$K^\Sigma$ is a compact subset of $\Sigma$. 
Note that, by the properties of the foliation,  
we have $cl(o_n)=cl(D(G_n))$ is contained in  $F(K^\Sigma,\mathbb{R})$ 
for any $n$. 
Consider now the cylinder  $F(K^\Sigma,[t_1,t_2])$. We claim 
that  there exists a $k$ such that 
$cl(o_n)\subseteq F(K^\Sigma,[t_1,t_2])$ for $n\geq k$. 
If it were not so, there should be  a sequence of points 
$x_n\in cl(o_n)\cap (F(K^\Sigma,t_1) \cup F(K^\Sigma,t_2))$.  
So, extract a subsequence $\{x_{s_n}\}$ which belongs say to 
$cl(o_{s_n})\cap F(K^\Sigma,t_1)$. This is a compact set because
so is  $F(K^\Sigma,t_1)$. So there is  a subsequence 
$\{x_{r_{s_n}}\}$ converging to a point $x$. 
Clearly $x\in F(K^\Sigma,t_1)$. Moreover 
$x\in cl(o)$ because  $x\in cl(o_{r_{s_n}})$ for any $n$. 
This leads to a contradiction because by construction 
$cl(o)\subset F(K^\Sigma,(t_1,t_2))$. 
Finally, for $n\geq k$ the  sets 
$o_n,a_n$ are contained in the compact set $F(K^\Sigma,[t_1,t_2])$,
so they are relatively compact, and the proof follows.
\end{proof}

\begin{lemma}
\label{Z:4}
Let $o$ be a diamond of $M$ and let $W$ be an open subset of 
$M$ such that $cl(o)\subset W$. Then there are two diamonds 
$\widetilde o$ and $a$ based on the same Cauchy surface as
$o$ such that  $cl(\widetilde o)\subset W$, 
$cl(o),cl(a)\subset \widetilde o$ and $cl(o)\perp
cl(a)$.
\end{lemma}
\begin{proof}
Let $\{o_n,a_n\}$ be a sequence of diamonds 
satisfying the properties of Lemma \ref{Z:3}. The idea of the proof 
is the same as  the second step of the proof of Lemma \ref{Z:3}. Briefly,
assume that $cl(o_n)\cap (M\setminus W)\ne\emptyset$ for any $n$. 
Take a sequence of points $\{x_n\}$ with $x_n\in cl(o_n)\cap (M\setminus W)$.
Since $cl(o_n)\cap (M\setminus W)\subset cl(o_1)$, for any $n$, 
and since the latter is compact, 
there is a subsequence $\{x_{s_n}\}$ converging to a point $x$. 
The limit point $x$ is in $cl(o)\cap (M\setminus W)$ and this lead
to a contradiction. So there is $k$ such that for $n\geq k$
$cl(o_n)\subset W$, and the proof follows by the properties 
of the sequence   $\{o_n,a_n\}$.   
\end{proof}

We now show a homotopy deformation result whose proof 
is very similar to the proof of the Van Kampen theorem 
about the fundamental groups of topological spaces (see \cite{Gra}).
\begin{lemma}
\label{Z:5}
Let $M$ be a globally hyperbolic spacetime with dimension 
$d\geq 3$. Let $o$ be a diamond of $M$. If 
$\gamma:[0,1]\to M$ is a curve with $\gamma(0),\gamma(1)\in
o^{\minperp}$, then $\gamma$ is homotopic to a curve $\tilde\gamma$ lying in 
$o^{\minperp}$. 
\end{lemma}
\begin{proof}
Let $x=\gamma(0)$ and $y=\gamma(1)$.
Since $o^\minperp$ is connected, there is a curve 
$\tilde \gamma:[0,1]\to o^\minperp$ such that 
$\tilde \gamma(0) = y$, $\tilde \gamma(1) = x$. 
So, the composition $\gamma_1*\gamma$ is a loop over  $x$.\smallskip

By the definition of causal complement and since diamonds
are relatively compact, we have that $o^\minperp = (cl(o))^\minperp$.
So the point $x$ is causally disjoint from $cl(o)$. Let $G$ be 
the base of the diamond $o$. By Lemma \ref{Z:4} 
there is a  diamond $o_1$ whose base $G_1$ is contained in the
same Cauchy surface as $G$ and such that $cl(G)\subset G_1$, 
$G_1\setminus cl(G)$ is connected and $cl(o_1)\perp x$. 
By \cite{BS06}, there is a Cauchy surface 
$\cC$ which contains $G,G_1$ and $x$ and $cl(G_1)$ is disjoint from 
$x$.  Since any Cauchy surface is a deformation
retract of $M$ the loop $\gamma_1*\gamma$ is homotopic 
to a loop $\gamma_2$ lying in $\cC$. The 
loop $\gamma_2$ meets $cl(G)$ (otherwise the proof
would be completed), hence it meets $G_1\setminus cl(G)$. Let 
$z$ be a point where $\gamma_1$ meets $G_1\setminus cl(G)$.
Since $\cC\setminus cl(G)$  is connected there is a curve $\tau$, in 
$\cC\setminus cl(G)$, from $x$ to $z$. Then
$\overline{\tau}*\gamma_2*\tau$
is a loop over $z$ ($\overline{\tau}$ is the reverse of $\tau$).\smallskip

Since $G_1$ and 
$\cC\setminus cl(G)$  form an open  
cover of $\cC$, we have that  
$\overline{\tau}*\gamma_1*\tau\sim \beta_n* \beta_{n-1}*\cdots *\beta_1$ where 
$\beta_i$ is a curve, in $\cC$, contained either in $G_1$ or 
$\cC\setminus cl(G)$ (the Lebesgue's covering lemma for $[0,1]$).
However we can assume that any $\beta_i$ is a loop, over $z$,
either in $G_1$ or $\cC\setminus cl(G)$, 
because $G_1\setminus cl(G)$ is connected. Hence observing that 
$G_1$ is contractible, we have that 
$\overline{\tau}*\gamma_1*\tau\sim \beta$, where  $\beta$ is a loop
over $z$ in $\cC\setminus cl(G)$. Thus 
$\gamma_2\sim \tau*\beta*\overline{\tau}$.
Clearly $\tau*\beta*\overline{\tau}$ lies in $o^\minperp$.
Since $\gamma_2\sim \gamma_1*\gamma$, then 
$\gamma\sim \tilde\gamma$ where 
$\tilde\gamma\doteq\overline{\gamma_1}*\tau*\beta*\overline{\tau}$ is
a curve from $x$ to $y$ lying in $o^\minperp$. 
\end{proof} 
It is obvious but worth observing that the 
same result holds if we replace in the statement of this lemma 
$o^\minperp$ with the causal 
complement $x^\minperp$ of a point $x$ of $M$.\smallskip

We now provide a version of this lemma in terms of the poset
$K(M)$.
\begin{corollary}
\label{Z:6}
Let $M$ be a globally hyperbolic spacetime with dimension 
$d\geq 3$. Let $o$ be a diamond of $M$.  
\begin{itemize}
\item[(i)]  If $p$ is a path with $\partial_1p$, $\partial_0p$ 
in $o^\minperp$, then 
$p$ is homotopic to a path 
$q$ whose support  is contained in $o^\minperp$.
\item[(ii)] If $p$ is a path with 
$cl(\partial_1p)$, $cl(\partial_0p)$ 
in $o^\minperp$, then $p$ is homotopic to a path
$q$ whose support has closure  contained in $o^\minperp$.
\end{itemize}
\end{corollary}
\begin{proof}
The proof is an easy consequence of 
Lemma \ref{Z:5} and   \cite[Lemma 2.17]{Ruz3}. 
\end{proof}
As before we observe that the results of this corollary 
hold if we replace in the statement 
$o^\minperp$ with the causal complement $x^{\minperp}$ of 
a point $x$ of $M$.

\end{document}